\documentclass[12pt, draftclsnofoot, onecolumn]{IEEEtran}
\usepackage{algorithm}
\usepackage{graphicx}
\usepackage[noend]{algpseudocode}
\usepackage{etex} 
\usepackage{tikz}
\usetikzlibrary{shapes,arrows}
\usepackage{pstricks}
\usepackage{pst-node,pst-blur}
\usetikzlibrary{arrows}
\usepackage{amsfonts}
\usepackage{tabularx}
\usepackage{subfigure}
\usepackage{amssymb}
\usepackage{booktabs}
\usepackage{multirow}
\usepackage{epsfig}
\usepackage{epstopdf}
\usepackage{array}

\usepackage[noadjust]{cite}
\newtheorem{theorem}{Theorem}
\newtheorem{remark}{Remark}

\newtheorem{proposition}{Proposition}
\newtheorem{lemma}{Lemma}

\usepackage[cmex10]{amsmath}
\usepackage[cmex10]{mathtools}
\DeclarePairedDelimiter{\ceil}{\lceil}{\rceil}

\allowdisplaybreaks  
   
 \makeatletter 
 \def\@eqnnum{{\normalsize \normalcolor (\theequation)}} 
  \makeatother
\begin{document}
\title{Transmit Precoding and Receive Power Splitting for Harvested Power Maximization in MIMO SWIPT Systems}
\author{Deepak Mishra,~\IEEEmembership{Member,~IEEE} and George~C.~Alexandropoulos,~\IEEEmembership{Senior~Member,~IEEE}
\thanks{D. Mishra was with Department of Electrical Engineering, Indian Institute of Technology Delhi, 110016 New Delhi, India. Now he is with   Department of Electrical Engineering, Link\"oping University, 58183 Link\"oping, Sweden (e-mail: deepak.mishra@liu.se).}
\thanks{G. C. Alexandropoulos is with the Mathematical and Algorithmic Sciences Lab, Paris Research Center, Huawei Technologies France SASU, 92100 Boulogne-Billancourt, France (e-mail: george.alexandropoulos@huawei.com). 
}
\thanks{A preliminary conference version~\cite{CAMSAP17} of this work was presented at the IEEE Int. Wksp. on Comput. Adv. in Multi-Sensor Adaptive Process. (CAMSAP), Cura{\c c}ao, Dutch Antilles, Dec. 2017.}
}  

\maketitle

\begin{abstract}
We consider the problem of maximizing the harvested power in Multiple Input Multiple Output (MIMO) Simultaneous Wireless Information and Power Transfer (SWIPT) systems with power splitting reception. Different from recently proposed designs, with our optimization problem formulation we target for the jointly optimal transmit precoding and receive uniform power splitting (UPS) ratio maximizing the harvested power, while ensuring that the quality-of-service requirement of the MIMO link is satisfied. We assume practical Radio-Frequency (RF) energy harvesting (EH) receive operation that results in a non-convex optimization problem for the design parameters, which we first formulate in an equivalent generalized convex problem that we then solve optimally. We also derive the globally optimal transmit precoding design for ideal reception. Furthermore, we present analytical bounds for the key variables of both considered problems along with tight high signal-to-noise ratio approximations for their optimal solutions. Two algorithms for the efficient computation of the globally optimal designs are outlined. The first requires solving a small number of non-linear equations, while the second is based on a two-dimensional search having linear complexity. Computer simulation results are presented validating the proposed analysis, providing key insights on various system parameters, and investigating the achievable EH gains over benchmark schemes.
\end{abstract} 

\section{Introduction}\label{sec:introduction}  
There has been recently increasing interest~\cite{Niyato_Survey1,ComMag,SWIPT_modern} in utilizing Radio Frequency (RF) signals for transferring simultaneously energy and data, also known as Simultaneous Wireless Information and Power Transfer (SWIPT). {This technology has the potential to play a major role in the practical ubiquitous deployment of low power wireless devices in fifth generation (5G) wireless networks and beyond~\cite{SWIPT_modern,Rev1,Rev2,Rev3}. Particularly, the SWIPT technology in conjunction with the adoption of wireless devices capable of performing Energy Harvesting (EH) is one of the promising candidates for enabling the perpetual operation of small cells, Internet-of-Things (IoT)~\cite{Niyato_Survey1}, Machine-to-Machine (M2M) communications and cognitive radio networks~\cite{Rev1,Rev2,Rev3}.}

Although the SWIPT concept has been lately very attractive and considered promising for empowering future wireless devices, it suffers from some fundamental bottlenecks. First and foremost, the signal processing and resource allocation strategies for wireless information and energy transfer differ significantly for achieving their respective goals~\cite{Vash,SMT}. As a matter of fact, there exists a non-trivial trade off between information and energy transfer that necessitates thorough investigation for optimizing the SWIPT performance. In addition, this performance is impacted by the low energy sensitivity and RF-to-Direct Current (DC) rectification efficiency~\cite{ComMag}. Another practical problem with SWIPT is the fact that the existing RF EH circuits cannot decode the information directly and vice-versa~\cite{WCL18,Tradeoff}. Lastly, the available solutions~\cite{MIMO_SWIPT,TCOM2} for realizing practically achievable SWIPT gains require high complexity and are still far from providing analytical insights on the optimum SWIPT performance. To confront with the latter bottlenecks, the Multiple-Input-Multiple-Output (MIMO) technology and resource allocation schemes as well as cooperative relaying strategies have been recently considered~\cite{WCL18,ComMag,DPS,MIMO_SWIPT,Tradeoff,RP,Spatial,MISO-PS,SOCP,MUTxMIMO,ASUPS,FDPowerMn,TCOM2,Close}. In this paper, we are interested in MIMO communication systems that offer spatial degrees of freedom which can be used for SWIPT, and we next discuss the relevant literature.

\subsection{State-of-the-Art}\label{sec:review} 
The non-trivial trade off between information capacity and average received power {for EH} was firstly investigated in the pioneering works~\cite{Vash,SMT} for a Single-Input-Single-Output (SISO) link operating over both frequency selective and non-selective channels corrupted by Additive White Gaussian Noise (AWGN). Then, the authors in~\cite{Tradeoff} discussed why the SWIPT theoretical gains are difficult to realize in practice and proposed some practical Receiver (RX) {architectures}. Among them belong the Time Switching (TS), Power Splitting (PS), and Antenna Switching (AS)~\cite{DPS} architectures that use one portion of the received signal (in time, power, or space) for EH and another one for Information Decoding (ID). In~\cite{MIMO_SWIPT}, Transmitter (TX) precoding techniques for efficient SWIPT in RF-powered MIMO systems were presented. Recently, the Spatial Switching (SS) was proposed~\cite{Spatial} that first decomposes the MIMO channel to its spatial eigenchannels and then assigns some for energy and some for information transfer~\cite{WCL18}.

The aforementioned RX architectures for SWIPT have been lately considered in various MIMO system setups~\cite{Spatial,MISO-PS,SOCP,MUTxMIMO,ASUPS,FDPowerMn,Close}. For example, the transmit power minimization satisfying both energy and rate requirements was investigated in \cite{Spatial} for MIMO SWIPT systems with SS. In \cite{MISO-PS}, a Semi-Definite Programming (SDP) relaxation technique for a multi-user multiple-input single-output system was used to study the joint TX precoding and PS optimization for minimizing the transmit power under signal-to-interference-plus-noise ratio and EH constraints. A second-order cone programming relaxation solution for the latter problem with significantly reduced computational complexity than SDP was proposed in \cite{SOCP}. In~\cite{MUTxMIMO} and~\cite{ASUPS}, more general MIMO interference channels were investigated adopting the interference alignment technique. The authors in~\cite{FDPowerMn} considered a multi-antenna full duplex access point and a single-antenna full duplex user, and investigated the joint design of TX precoding and RX PS ratio for minimizing the weighted sum transmit power. However, the vast majority of the available MIMO SWIPT works presented suboptimal iterative algorithms based on convex relaxation and approximation approaches that are unable to provide key insights on the optimal TX precoding and PS design.

\subsection{Motivation and Contribution}\label{sec:motiv} 
A major goal of RF EH systems is the optimization of the end-to-end EH efficiency~\cite{Niyato_Survey1} by maximizing the rate-constrained harvested energy for a given TX power budget. This is in principle challenging with the available EH circuitry implementations, where the RF-to-DC rectification is a non-linear function of the received RF power~\cite{NonRFH,Powercast,TCOM,Close}. This fact leads naturally to the necessity of optimizing the harvested power rather than the receiver power treated in the existing literature~\cite{MIMO_SWIPT,DPS,RP,Spatial,MISO-PS,SOCP,MUTxMIMO,ASUPS,FDPowerMn,TCOM2}; therein, constant RF-to-DC rectification efficiency has been assumed. 
In this paper, we study the problem of maximizing the harvested power in MIMO SWIPT systems with practical PS reception~\cite{MIMO_SWIPT}, while ensuring that the quality-of-service requirement of the MIMO link is met. We note that, although the PS architecture involves higher RX complexity, it is more efficient than TS since the received signal is used for both EH and ID. In addition, PS is more suitable for delay-constraint applications. We are interested in finding the jointly optimal TX precoding scheme and the RX Uniform PS (UPS) ratio for the considered optimization problem, and in gaining analytical insights on the interplay among various system parameters. \textcolor{black}{To our best of knowledge, this joint optimization problem for maximizing the harvested DC power has not been considered in the past, and available designs for practical MIMO SWIPT are suboptimal. In addition, although \cite{Close} considered non-linear RF EH modeling for investigating the SWIPT rate-energy tradeoff, analytical insights on the joint globally optimal design and efficient algorithmic implementations to obtain it were missing.} 

The key contributions of this paper are summarized as follows. 
\begin{itemize}
\item \textcolor{black}{We present an equivalent generalized convex formulation for the considered non-convex harvested power maximization problem that helps us in deriving the global jointly optimal TX precoding and RX UPS ratio design. We also present the globally optimal TX precoding design for ideal reception. For both designs there exists a rate requirement value that determines whether the TX precoding operation is energy beamforming or information spatial multiplexing. This \textit{novel} feature stems from our novel problem formulation involving rate constrained EH optimization and does not appear in  available designs~\cite{MIMO_SWIPT,Spatial,MISO-PS,SOCP,MUTxMIMO,ASUPS,FDPowerMn,Close}.}   
\item We investigate the trade off between the harvested power and achievable information rate for both globally optimal designs. Practically motivated asymptotic analysis for obtaining globally optimal solutions in the high Signal-to-Noise-Ratio (SNR) regime in a computationally efficient manner is also provided. 
\item We detail a computationally efficient algorithm for the global jointly optimal design and present a low complexity alternative algorithm that is based on a two-dimensional (2-D) linear search. The complexity of the latter algorithm is \textit{linear} in the number of the spatial eigenchannels of the MIMO system. Both algorithms can also be straightforwardly used for implementing the globally optimal TX precoding design for ideal reception.
\item We carry out a detailed numerical investigation of the presented optimal solutions to provide insights on the interplay among various system parameters on the trade off between harvested power and achievable information rate.
\end{itemize}
\textcolor{black}{The key challenges with our problem formulation addressed in this paper include its generalized convexity proof given the non-linear rectification property and the analytical exploration of non-trivial insights on its controlling variables, which helped us in designing a low complexity global optimization algorithm. Additionally, we would like to emphasize that our performance results are valid for any practical RF EH circuit model~\cite{NonRFH,Powercast,Close}, and our key insights on the optimal transceiver design parameters can be extended to investigate multiuser MIMO SWIPT systems.}

\subsection{Paper Organization and Notations}\label{sec:OrgNot} 
The considered system model is described in Section~\ref{sec:system_model}, while Section~\ref{sec:problem} introduces the joint

\noindent optimization framework. Section~\ref{sec:soln} includes the globally optimal solutions, and analytical bounds and approximations are presented in Section~\ref{sec:ana}. A detailed numerical investigation of the proposed joint design is provided in Section~\ref{sec:results}, whereas Section~\ref{sec:conclusion} concludes the paper.

Vectors and matrices are denoted by boldface lowercase and boldface capital letters, respectively. The transpose and Hermitian transpose of $\mathbf{A}$ are denoted by $\mathbf{A}^{\rm T}$ and $\mathbf{A}^{\rm H}$, respectively, and $\det(\mathbf{A})$ is the determinant of $\mathbf{A}$, while $\mathbf{I}_{n}$ ($n\geq2$) is the $n\times n$ identity matrix and $\mathbf{0}_{n}$ ($n\geq2$) is the $n$-element zero vector. The trace of $\mathbf{A}$ is denoted by $\mathrm{tr}\left(\mathbf{A}\right)$, $[\mathbf{A}]_{i,j}$ stands for $\mathbf{A}$'s $(i,j)$-th element, $\lambda_{\max}\left(\mathbf{A}\right)$ represents the largest eigenvalue of $\mathbf{A}$, and ${\rm diag}\{\cdot\}$ denotes a square diagonal matrix with $\mathbf{a}$'s elements in its main diagonal. $\mathbf{A}^{-1}$ and $\mathbf{A}^{1/2}$ represent the inverse and square-root, respectively, of a square matrix $\mathbf{A}$, whereas $\mathbf{A}\succeq0$ and $\mathbf{A}\succ0$ mean that $\mathbf{A}$ is positive semi definite and positive definite, respectively. $\mathbb{C}$ represents the complex number set, $\left(x\right)^+\triangleq\max\left\{0,x\right\}$, $\ceil{x}$ denotes the smallest integer larger than or equal to $x$, $\mathbb{E}\{\cdot\}$ denotes the expectation operator, and $\mathcal{O}\left(\cdot\right)$ is the Big O notation \cite[p$.$ 517]{Baz} denoting order of complexity. 

\begin{figure}[!t]
\centering
{{\includegraphics[width=4.2in]{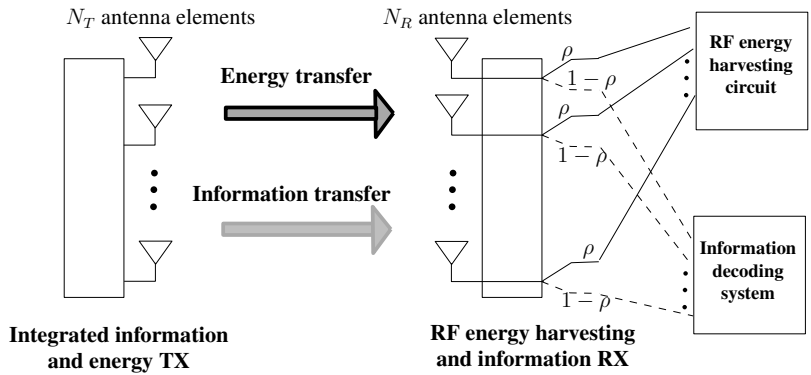} }}
\caption{\small MIMO SWIPT system with UPS reception. $\rho$ denotes the common PS ratio at each receive antenna element.}
    \label{fig:model}
\end{figure}
 
\section{System and Channel Models}\label{sec:system_model} 
We consider the MIMO SWIPT system of Fig.~\ref{fig:model}, where the TX is equipped with $N_T$ antenna elements and wishes to simultaneously transmit information and energy to the RF-powered RX having $N_R$ antenna elements. We assume a frequency flat MIMO fading channel $\mathbf{H}\in\mathbb{C}^{N_R\times N_T}$ that remains constant during one transmission time slot and changes independently from one slot to the next. The channel is assumed to be perfectly known at both TX and RX. The entries of $\mathbf{H}$ are assumed to include independent, zero-mean circularly symmetric complex Gaussian (ZMCSCG) random variables with unit variance; this assumption ensures that the rank of $\mathbf{H}$ is given by $r=\min(N_R,N_T)$. The baseband received signal $\mathbf{y}\in\mathbb{C}^{N_R\times 1}$ at RX is given by
\begin{equation}\label{eq:sys_model}
\mathbf{y} = \mathbf{H}\mathbf{x} + \mathbf{n},
\end{equation}
where $\mathbf{x}\in\mathbb{C}^{N_T\times 1}$ denotes the transmitted signal with covariance matrix $\mathbf{S}\triangleq\mathbb{E}\{\mathbf{x}\mathbf{x}^{\rm H}\}$ and $\mathbf{n}\in\mathbb{C}^{N_R\times 1}$ represents the AWGN vector having ZMCSCG entries each with variance $\sigma^2$. The elements of $\mathbf{x}$ are assumed to be statistically independent, the same is assumed for the elements of $\mathbf{n}$. We also make the usual assumption that the signal elements are statistically independent with the noise elements. For the transmitted signal we finally assume that there exists an average power constraint across all TX antennas denoted by $\mathrm{tr}\left(\mathbf{S}\right)\le P_T$.

\begin{figure}[!t]
	\centering
	\subfigure[{Powercast P1110 EVB characteristics~\cite{Powercast}.}]{\makebox[0.4\linewidth]{\includegraphics[width=2.25in]{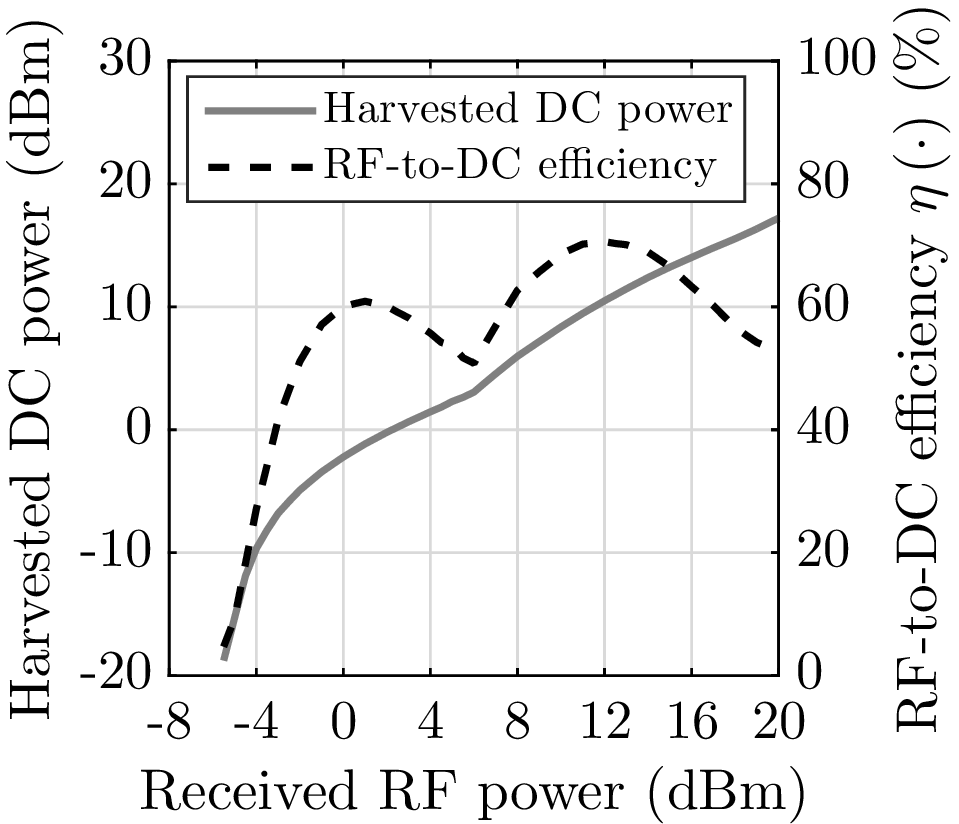} }}\qquad\qquad
	\subfigure[{Characteristics of EH circuit designed in~\cite{RFEH2008}.}]
	{\makebox[0.4\linewidth]{\includegraphics[width=2.25in]{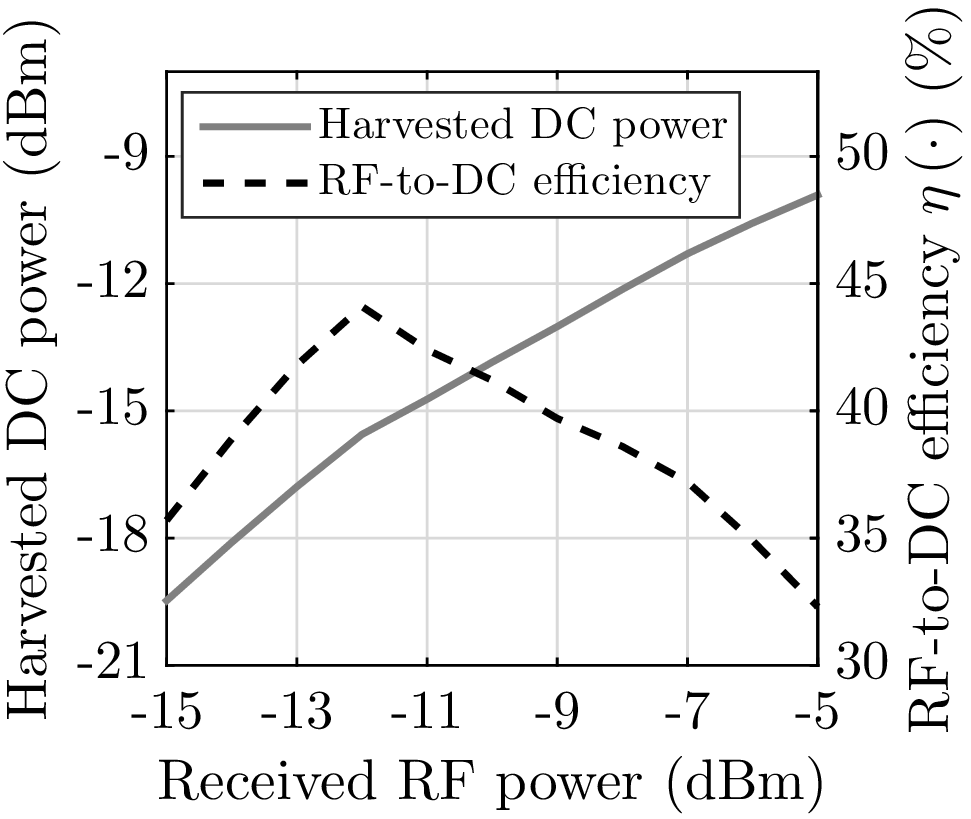} }}
	\caption{\small 
		{Variation of harvested DC power and RF-to-DC efficiency  with received RF power for practical two circuits.}} 
	\label{fig:NonLinear-RFEH} 
\end{figure}

Capitalizing on the signal model in \eqref{eq:sys_model}, the {average} received power $P_R$ across all RX antennas can be obtained as $P_R\triangleq\mathbb{E}\{\mathbf{y}^{\rm H}\mathbf{y}\}$. Note that the averaging is performed over the transmitted symbols during each coherent channel block. {As} the noise strength (generally lower than $-80$dBm) is much below than the received energy sensitivity of practical RF {EH} circuits (which is around $-20$dBm)~\cite{Niyato_Survey1}, we next neglect the contribution of $\mathbf{n}$ to the harvested power. Note, however, that the analysis and optimization results of this paper can be easily extended for non-negligible noise power scenarios. We therefore rewrite $P_R$ as the following function of $\mathbf{H}$ and $\mathbf{x}$ 
\begin{equation}\label{eq:PR}
P_R\triangleq\mathbb{E}\left\{\mathbf{x}^{\rm H}\mathbf{H}^{\rm H}\mathbf{H}\mathbf{x}\right\}=\mathrm{tr}\left(\mathbf{H}\mathbf{S}\mathbf{H}^{\rm H}\right).
\end{equation} 
As demonstrated in Fig.~\ref{fig:model}, we consider the UPS ratio $\rho\in[0,1]$ at each RX antenna element. This ratio reveals that $\rho$ fraction of the received signal power at each antenna is used for RF EH, while the remaining $1-\rho$ fraction is used for ID. With this setting together with the previous noise assumption, the average total received power $P_{R,E}$ available for RF EH is given by $P_{R,E}\triangleq\rho P_R=\rho\,\mathrm{tr}\left(\mathbf{H}\mathbf{S}\mathbf{H}^{\rm H}\right)$. This definition for the average received power is the most widely used definition~\cite{MIMO_SWIPT,TCOM2} for investigating the performance lower bound with the PS RX architectures. Supposing that $\eta\left(\cdot\right)$ denotes the RF-to-DC rectification efficiency function, which is in general a non-linear positive function of the received RF power $P_{R,E}$ available for EH~\cite{Powercast,NonRFH,Close}, the total harvested DC power is obtained as $P_H\triangleq\eta\left(\rho P_R\right)\rho P_R$. Despite this circuit dependent non-linear relationship between $\eta$ and $P_{R,E}$, we note that $P_H$ is monotonically non-decreasing in $P_{R,E}=\rho P_R$ for any practical RF EH circuit~\cite{NonRFH,Powercast,Close} due to the law of energy conservation. {For instance, to give more insights, we plot both  $\eta$ and $P_H\triangleq\eta\left(P_{R,E}\right)\,P_{R,E}$ as a function of the received RF power $P_{R,E}$ variable at the input of two real-world RF EH circuits, namely, (i) the commercially available Powercast P1110 evaluation board (EVB)~\cite{Powercast} and (ii) the circuit designed in~\cite{RFEH2008} for low power far field RF EH in Figs.~\ref{fig:NonLinear-RFEH}(a) and~\ref{fig:NonLinear-RFEH}(b), respectively. So, using the relationship  $P_{H}=\mathcal{F}\left(P_{R,E}\right)$, where $\mathcal{F}\left(\cdot\right)$ represents a \textit{non-linear non-decreasing} function, we are able to obtain the jointly global optimal design.}


\section{Joint TX and RX Optimization Framework}\label{sec:problem}
In this section, we present the mathematical formulation of the optimization problem under investigation. We first consider in Section~\ref{sec:formulate} the practical case of UPS reception and prove an interesting property of the underlying optimization problem that will be further exploited in Section~\ref{sec:soln} for deriving the globally optimal design for the unknown system parameters. Aiming at comparing with the ideal reception case, we present in Section~\ref{sec:ideal} a mathematical formulation including only the TX precoding design.

\subsection{UPS Reception}\label{sec:formulate}
Focusing on the MIMO SWIPT model of Section~\ref{sec:system_model}, we consider the problem of designing the covariance matrix $\mathbf{S}$ at the multi-antenna TX and the UPS ratio $\rho$ at the multi-antenna RF EH RX for maximizing the total harvested DC power, while satisfying a minimum instantaneous rate requirement ${R}$ in bits per second per Hz (bps/Hz) for information transmission. {We have adopted UPS because it not only helps in attaining global-optimality of the proposed joint design, but also leads to  an efficient low complexity algorithmic implementation in Section~\ref{sec:2D}.} Our proposed design framework can be mathematically expressed by the following optimization problem: 
\begin{equation*}\label{eqOPT}
\begin{split}
  \mathcal{OP}: &\max_{\rho,\mathbf{S}} \quad P_H=\eta\left(\rho\,\mathrm{tr}\left(\mathbf{H}\mathbf{S}\mathbf{H}^{\rm H}\right)\right)\rho\,\mathrm{tr}\left(\mathbf{H}\mathbf{S}\mathbf{H}^{\rm H}\right)
	\\& \hspace{0.6cm}\textrm{s.t.}~~({\rm C1}):~\log_2\left(\det\left(\mathbf{I}_{N_R}+\left(1-\rho\right)\sigma^{-2}\mathbf{H}\mathbf{S}\mathbf{H}^{\rm H}\right)\right)\ge{R}, 
	\\& \hspace{1.39cm}               ({\rm C2}):~\mathrm{tr}\left(\mathbf{S}\right)\leq P_T,~({\rm C3}):~\mathbf{S}\succeq 0,~({\rm C4}):~0\leq\rho\leq1,
\end{split}
\end{equation*}
where constraint $({\rm C1})$ represents the minimum instantaneous rate requirement, $({\rm C2})$ is the average transmit power constraint, while constraints $({\rm C3})$ and $({\rm C4})$ are the boundary conditions for $\mathbf{S}$ and $\rho$. It can be easily concluded from $\mathcal{OP}$ that the objective function $P_H$ is jointly non-concave in regards to the unknown variables $\mathbf{S}$ and $\rho$. It will be shown, however, in the following Lemma~\ref{lem:pcc} that the received power $P_{R,E}$ available for EH is jointly pseudoconcave in $\mathbf{S}$ and $\rho$.  

\begin{lemma}\label{lem:pcc}
The RF received power $P_{R,E}$ is a joint pseudoconcave function of $\mathbf{S}$ and $\rho$.
\end{lemma} 
\begin{IEEEproof}
With $\mathrm{tr}\left(\mathbf{H}\mathbf{S}\mathbf{H}^{\rm H}\right)$ being linear in $\mathbf{S}$, we deduce that the total average received RF power $P_{R,E}=\rho\,\mathrm{tr}\left(\mathbf{H}\mathbf{S}\mathbf{H}^{\rm H}\right)$ available for EH is the product of two positive linear (or concave) functions of $\rho$ and $\mathbf{S}$. Since the product of two positive concave functions is log-concave~\cite[Chapter 3.5.2]{boyd} and a positive log-concave function is also pseudoconcave~\cite[Lemma 5]{TCOM2}, the joint pseudoconcavity of $P_{R,E}$ with respect to $\mathbf{S}$ and $\rho$ is proved.
\end{IEEEproof}

We now show that solving $\mathcal{OP}$ is equivalent to solving the following optimization problem:
\begin{equation*}\label{eqOPT1}
  \mathcal{OP}1: \max_{\rho,\mathbf{S}} \quad P_{R,E}=\rho\,\mathrm{tr}\left(\mathbf{H}\mathbf{S}\mathbf{H}^{\rm H}\right)~~
	\textrm{s.t.}~~({\rm C1}), ({\rm C2}), ({\rm C3}), ({\rm C4}). 
\end{equation*}
\begin{proposition}\label{prep}
The solution pair $\left(\mathbf{S}^*,\rho^*\right)$ of $\mathcal{OP}1$ solves $\mathcal{OP}$ optimally.
\end{proposition} 
\begin{IEEEproof}
Irrespective of the circuit-dependent non-linear relationship between $\eta$ and $P_{R,E}$, $P_H$ is always monotonically non-decreasing in $P_{R,E}$~\cite{NonRFH,Powercast,Close}. It can be concluded from~\cite{avriel2010generalized,boyd} that the monotonic non-decreasing transformation $P_H$ of the pseudoconcave function $P_{R,E}$ is also pseudoconcave and possesses the unique global optimality property~\cite[Props$.$ 3.8 and 3.27]{avriel2010generalized}. This reveals that $\mathcal{OP}$ and $\mathcal{OP}1$ are equivalent~\cite{Baz}{,} sharing the same solution pair $\left(\mathbf{S}^*,\rho^*\right)$.    
\end{IEEEproof}
It can be deduced from Proposition~\ref{prep} that one may solve $\mathcal{OP}1$ and then use the resulting maximum received power $P_{R,E}^*=\rho^*\,\mathrm{tr}\left(\mathbf{H}\mathbf{S}^*\mathbf{H}^{\rm H}\right)$ to compute the maximum harvested DC power as $P_H^*=\eta\left(P_{R,E}^*\right)P_{R,E}^*$. Although $\mathcal{OP}1$ is a non-convex problem, we prove in the following theorem a specific property for it that will {be used} in Section~\ref{sec:soln} to derive its optimal solution. 
\begin{theorem}\label{th:Gen_Cvx}
$\mathcal{OP}1$ is a generalized convex problem and its globally optimal solution $\left(\mathbf{S}^*,\rho^*\right)$ can be obtained by solving its Karush-Kuhn-Tucker (KKT) conditions.
\end{theorem} 
\begin{IEEEproof}
As shown in Lemma~\ref{lem:pcc}, $P_{R,E}$ is a joint pseudoconcave function of $\mathbf{S}$ and $\rho$. It follows from constraint $({\rm C1})$ that the function $R-\log_2\left(\det\left(\mathbf{I}_{N_R}+\left(1-\rho\right)\sigma^{-2}\mathbf{H}\mathbf{S}\mathbf{H}^{\rm H}\right)\right)$ is jointly convex on $\rho$ and $\mathbf{S}$; this ensues from the fact that the matrix inside the determinant is a positive definite matrix~\cite{MIMO_SWIPT,MISO-PS,SOCP,MUTxMIMO}. In addition, constraints $({\rm C2})$ and $({\rm C3})$ are linear with respect to $\mathbf{S}$ and independent of $\rho$, and constraint $({\rm C4})$ depends only on $\rho$ and is convex. The proof completes by combining the latter findings and using them in~\cite[Theorem 4.3.8]{Baz}.        
\end{IEEEproof}

\color{black}Capitalizing on the findings of Proposition~\ref{prep} and Theorem~\ref{th:Gen_Cvx}, we henceforth focus on the maximization of the received RF power $P_{R,E}$ for EH. The jointly optimal TX precoding and UPS design for this problem will also result in the maximization of the harvested DC power $P_H^*$ for any practical RF EH circuitry. We note that the proposed joint transceiver design in this paper is different from the ones in the existing works~\cite{MIMO_SWIPT,DPS,RP,Spatial,MISO-PS,SOCP,MUTxMIMO,ASUPS,FDPowerMn,TCOM2} that considered the received RF power for EH as a constraint and used a trivial linear RF EH model for their investigation.\color{black}

\subsection{Ideal Reception}\label{sec:ideal} 
To investigate the theoretical upper bound for $P_{R,E}$, we consider in this section an ideal RX architecture that is capable of using all received RF power for both EH and ID. In particular, we remove $\rho$ from $\mathcal{OP}1$ and $({\rm C1})$ and consider the following optimization problem: 

\noindent\begin{equation*}\label{eqOPT2}
\begin{split}
  \mathcal{OP}2: &\max_{\mathbf{S}} \; P_R=\mathrm{tr}\left(\mathbf{H}\mathbf{S}\mathbf{H}^{\rm H}\right),
	\quad\textrm{s.t.}~~({\rm C2}),~({\rm C3}),~({\rm C5})\!:\log_2\left(\det\left(\mathbf{I}_{N_R}+\sigma^{-2}\mathbf{H}\mathbf{S}\mathbf{H}^{\rm H}\right)\right)\ge{R}.
\end{split}
\end{equation*}
From the findings in the proof of Theorem~\ref{th:Gen_Cvx}, the objective function $P_R$ of $\mathcal{OP}2$ along with constraints $({\rm C2})$ and $({\rm C3})$ are linear in $\mathbf{S}$. In addition, $({\rm C5})$ is convex due to the concavity of the logarithm with respect to $\mathbf{S}$. Combining the latter facts yields that $\mathcal{OP}2$ is a convex problem, and hence, its optimal solution can be found using the Lagrangian dual method~\cite{boyd,Baz}.

\section{Optimal TX Precoding and RX Power Splitting}\label{sec:soln}
We first investigate the fundamental trade off between energy beamforming and information spatial multiplexing {in $\mathcal{OP}1$}. Then, we present the global jointly optimal TX precoding and RX UPS design for $\mathcal{OP}1$ as well as the globally optimal TX precoding design for $\mathcal{OP}2$. 

\subsection{Energy Beamforming versus Information Spatial Multiplexing}\label{sec:tradeoff}
Let us consider the reduced Singular Value Decomposition (SVD) of the MIMO channel matrix $\mathbf{H}=\mathbf{U}\boldsymbol{\Lambda}\mathbf{V}^{\rm H}$, where $\mathbf{V}\in\mathbb{C}^{N_T\times r}$ and $\mathbf{U}\in\mathbb{C}^{N_R\times r}$ are unitary matrices and $\boldsymbol{\Lambda}\in\mathbb{C}^{r\times r}$ is the diagonal matrix consisting of the $r$ non-zero eigenvalues of $\mathbf{H}$ in decreasing order of magnitude. Ignoring the rate constraint $({\rm C1})$ in $\mathcal{OP}1$ (or equivalently in $\mathcal{OP}$) leads to the rank-$1$ optimal TX covariance matrix $\mathbf{S^*}=\mathbf{S_{_\mathrm{EB}}}\triangleq P_T\,\mathbf{v}_1\mathbf{v}_1^{\rm H}$~\cite{MIMO_SWIPT,MIMO_book}, where $\mathbf{v}_1\in\mathbb{C}^{N_T\times 1}$ is the first column of $\mathbf{V}$ that corresponds to the eigenvalue $[\boldsymbol{\Lambda}]_{1,1}\triangleq\sqrt{\lambda_{\max}\left(\mathbf{H}^{\rm H}\mathbf{H}\right)}$. This TX precoding, also known as \textit{transmit energy beamforming}, allocates $P_T$ to the strongest eigenmode of $\mathbf{H}^{\rm H}\mathbf{H}$ and is known to maximize the harvested or received power. On the other hand, it is also well known \cite{MIMO_WF} that one may profit from the existence of multiple antennas and channel estimation techniques to realize spatial multiplexing of multiple data streams, thus optimizing the information communication rate. Spatial multiplexing adopts the waterfilling technique to perform optimal allocation of $P_T$ over all the available eigenchannels of the MIMO channel matrix. Evidently, for our problem formulation $\mathcal{OP}1$ that includes the rate constraint $({\rm C1})$ and PS reception, we need to investigate the underlying fundamental trade off between TX energy beamforming and information spatial multiplexing. As previously described, these two transmission schemes have contradictory objectives, and thus provide different TX designs.

Suppose we adopt energy beamforming in $\mathcal{OP}1$, resulting in the received RF power $P_{R_{_{\mathrm{EB}}}}\triangleq\rho_{_\mathrm{EB}}P_T\,[\boldsymbol{\Lambda}]_{1,1}^2$ where $\rho_{_\mathrm{EB}}$ represents the unknown UPS parameter. To find the optimal UPS parameter $\rho_{_\mathrm{EB}}^*$, we need to seek for the best power allocation $(1-\rho_{_\mathrm{EB}}^*)$ for ID meeting the rate requirement $R$. To do so, we solve constraint $({\rm C1})$ at equality over the UPS parameter yielding   
\begin{align}\label{eq:reb}
\rho_{_\mathrm{EB}}^* \triangleq &\max\left\lbrace0,1-\frac{\left(2^{{R}}-1\right) \sigma ^2}{P_T\,[\boldsymbol{\Lambda}]_{1,1}^2}\right\rbrace. 
\end{align} 
It can be concluded that both $\rho_{_\mathrm{EB}}^*$ and the maximum received RF power given by $\rho_{_\mathrm{EB}}^*P_T\,[\boldsymbol{\Lambda}]_{1,1}^2$ are decreasing functions of ${R}$. This reveals that there exists a rate threshold $R_{\rm th}$ such that, when $R>R_{\rm th}$, one should allocate $P_T$ over to at least two eigenchannels instead of performing energy beamforming, i$.$e$.$, instead of assigning $P_T$ solely to the strongest eigenchannel. We are henceforth interested in finding this $R_{\rm th}$ value. Consider the optimum power allocation $p_1^*$ and $p_2^*$ {for} the two highest gained eigenchannels with eigenmodes $[\boldsymbol{\Lambda}]_{1,1}$ and $[\boldsymbol{\Lambda}]_{2,2}$, respectively, with $[\boldsymbol{\Lambda}]_{1,1}>[\boldsymbol{\Lambda}]_{2,2}$. By substituting these values into $({\rm C1})$ and solving at equality for the optimum UPS parameter $\rho_{_{\mathrm{SM}_2}}^*$ for spatial multiplexing over two eigenchannels deduces to  
\begin{eqnarray}\label{eq:rhoSM2}
&\rho_{_{\mathrm{SM}_2}}^*\triangleq 1+\frac{\sigma ^2}{2}\left(\frac{1}{[\boldsymbol{\Lambda}]_{1,1}^2 p_1^*}+\frac{1}{[\boldsymbol{\Lambda}]_{2,2}^2 p_2^*} +\frac{\sqrt{\left([\boldsymbol{\Lambda}]_{1,1}^2 p_1^*-[\boldsymbol{\Lambda}]_{2,2}^2 p_2^*\right)^2+2^{{R}+2}[\boldsymbol{\Lambda}]_{1,1}^2 [\boldsymbol{\Lambda}]_{2,2}^2 p_1^* p_2^* }}{[\boldsymbol{\Lambda}]_{1,1}^2 [\boldsymbol{\Lambda}]_{2,2}^2 p_1^* p_2^*}\right),
\end{eqnarray}
resulting in the maximum received RF power for EH given by $\rho_{_{\mathrm{SM}_2}}^*\left(p_1^*\,[\boldsymbol{\Lambda}]_{1,1}^2+p_2^*\,[\boldsymbol{\Lambda}]_{2,2}^2\right)$. We now combine the latterly obtained maximum received RF power with spatial multiplexing and that of energy beamforming to compute $R_{\rm th}$. The rate threshold value that renders energy beamforming more beneficial than spatial multiplexing in terms of received RF power can be obtained from the solution of the following inequality  
\begin{equation}\label{eq:inequality}
\rho_{_\mathrm{EB}}^*P_T[\boldsymbol{\Lambda}]_{1,1}^2>\rho_{_{\mathrm{SM}_2}}^*\left(p_1^*\,[\boldsymbol{\Lambda}]_{1,1}^2+p_2^*\,[\boldsymbol{\Lambda}]_{2,2}^2\right).
\end{equation}
Substituting \eqref{eq:reb} and \eqref{eq:rhoSM2} into \eqref{eq:inequality} and applying some algebraic manipulations yields
\begin{eqnarray}\label{eq:rth}
\textstyle R_{\mathrm{th}}\, \triangleq\,  \log_2\left(1+\frac{p_2^*([\boldsymbol{\Lambda}]_{1,1}^2-[\boldsymbol{\Lambda}]_{2,2}^2)}{\sigma ^2}+\sqrt{\frac{([\boldsymbol{\Lambda}]_{1,1}^2-[\boldsymbol{\Lambda}]_{2,2}^2) ([\boldsymbol{\Lambda}]_{1,1}^2 p_1^*+p_2^*[\boldsymbol{\Lambda}]_{2,2}^2)^2}{[\boldsymbol{\Lambda}]_{1,1}^2 [\boldsymbol{\Lambda}]_{2,2}^2 \sigma ^2 p_1^*}}\right).
\end{eqnarray}

\begin{figure}[!t]
\centering
{{\includegraphics[width=3.6in]{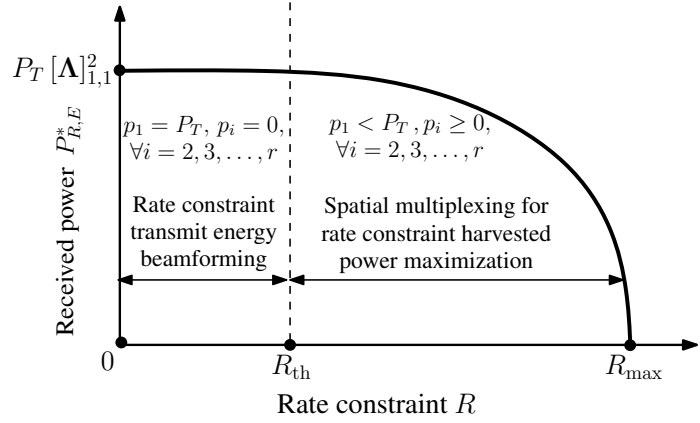} }}
\caption{\small The trade off between received power for EH and achievable information rate. The rate threshold $R_{\mathrm{th}}$ is the switching point between the transmit precoding modes: energy beamforming and information spatial multiplexing.}
    \label{fig:concept}
\end{figure}
\color{black}\begin{remark}
{The rate threshold $R_{\mathrm{th}}$ given by \eqref{eq:rth} evinces a switching point on the desired TX precoding operation, which is graphically presented in Fig$.$~\ref{fig:concept}. When the rate requirement $R$ is less or equal to $R_{\mathrm{th}}$, energy beamforming is sufficient to meet $R$, and hence, can be used for maximizing the received RF power. For cases where ${R}>{R_{\mathrm{th}}}$, statistical multiplexing needs to be adopted for maximizing the received RF power for EH while satisfying $R$. It is noted that this explicit non-trivial switching point $R_{\mathrm{th}}$ for the TX precoding mode is unique to the problem formulation considered in this paper, and has not been explored or investigated in the relevant literature~\cite{MIMO_SWIPT,Spatial,MISO-PS,SOCP,MUTxMIMO,ASUPS,FDPowerMn,Close} for the complementary problem formulations therein (i$.$e$.$, rate maximization or transmit power minimization subject to energy demands).}
\end{remark}\color{black}

{We next use the $R_{\mathrm{th}}$ definition given in \eqref{eq:rth} to obtain the global jointly optimal TX precoding and RX UPS design for $\mathcal{OP}1$ as well as the globally optimal TX precoding design for $\mathcal{OP}2$.}

\subsection{Globally Optimal Solution of $\mathcal{OP}1$}\label{sec:KKT}
Associating Lagrange multipliers $\mu$ and $\nu$ with constraints $({\rm C1})$ and $({\rm C2})$, respectively, while {keeping} $({\rm C3})$ and $({\rm C4})$ implicit, the Lagrangian function of $\mathcal{OP}1$ can be formulated as
\begin{eqnarray}\label{eq:Lang}
\textstyle\mathcal{L}\left(\mathbf{S},\rho,\mu,\nu\right) =\rho\,\mathrm{tr}\left(\mathbf{HSH}^{\rm H}\right)-\nu\left(\mathrm{tr}\left(\mathbf{S}\right)-P_T\right) -\mu\left({R}-\log_2\left[\det\left(\mathbf{I}_{N_R}+\frac{\left(1-\rho\right)\mathbf{H}\mathbf{S}\mathbf{H}^{\rm H}}{\sigma^{2}}\right)\right]\right).
\end{eqnarray}

Using Theorem~\ref{th:Gen_Cvx}, the globally optimal solution $\left(\mathbf{S^*},\rho^*\right)$ for $\mathcal{OP}1$ is obtained from the following four KKT conditions (subgradient and complimentary slackness conditions are defined, whereas the primal feasibility $({\rm C1})$--$({\rm C4})$ and dual feasibility constraints $\mu,\nu\ge 0$ are kept implicit):
\begin{subequations}\label{eq:KKT}
\begin{align}\label{eq:KKT1}
\frac{\partial \mathcal{L}}{\partial \mathbf{S}}=\frac{\mu\,\left(1-\rho\right)}{\sigma^2\,\ln2} \;\mathbf{H}^{\rm H}\left(\mathbf{I}_{N_R}+\left(1-\rho\right)\sigma^{-2}\mathbf{H}\mathbf{S}\mathbf{H}^{\rm H}\right)^{-1}\mathbf{H} +\rho\,\mathbf{H}^{\rm H}\mathbf{H}-\nu\mathbf{I}_{N_T}=0,
\end{align}
\begin{align}\label{eq:KKT2}
\frac{\partial \mathcal{L}}{\partial \rho}=-\mu\;\mathrm{tr}\left(\frac{\mathbf{HSH}^{\rm H}}{\sigma^2\,\ln2} \left(\mathbf{I}_{N_R}+\left(1-\rho\right)\sigma^{-2}\mathbf{H}\mathbf{S}\mathbf{H}^{\rm H}\right)^{-1}\right) +\mathrm{tr}\left(\mathbf{HSH}^{\rm H}\right)=0,
\end{align}
\begin{equation}\label{eq:KKT3}
\mu\left({R}-\log_2\left[\det\left(\mathbf{I}_{N_R}+\left(1-\rho\right)\sigma^{-2}\mathbf{H}\mathbf{S}\mathbf{H}^{\rm H}\right)\right]\right)=0,
\end{equation}
\begin{equation}\label{eq:KKT4}\nu\left(\mathrm{tr}\left(\mathbf{S}\right)-P_T\right)=0.
\end{equation}
\end{subequations} 
Solving the four equations included in \eqref{eq:KKT} yields the KKT point~\cite{Baz,boyd} defined by the optimal solution $\left(\mathbf{S^*},\rho^*,\mu^*,\nu^*\right)$. It is noted that it must hold $\nu\neq0$, because the total available transmit power $P_T$ is always fully utilized due to the monotonically increasing nature of the objective function $P_{R,E}$ in $\mathbf{S}$. This implies that $\mathrm{tr}\left(\mathbf{S}\right)= P_T$, i$.$e$.$, the sum of the power allocation is $P_T$, which means that constraint $({\rm C2})$ is always satisfied at equality. Similarly, it must hold $\mu\neq0$, because the received RF power is strictly increasing in $\rho$ and, as such, the remaining fraction $1-\rho$ allocated for ID needs to be sufficient in meeting  rate constraint $R$ that appears in {$({\rm C1})$}. 
 
Recalling the trade off discussion in Section~\ref{sec:tradeoff}, when $R\le{R_{\mathrm{th}}}$, the optimal TX covariance matrix is given as $\mathbf{S^*}=\mathbf{S_{_\mathrm{EB}}}$. For this case the optimum TX precoding operation is energy beamforming, i$.$e$.$, $\mathbf{F}\triangleq \mathbf{V}({\mathbf{P}^*})^{1/2}\in\mathbb{C}^{N_T\times r}$ where the $r\times r$ matrix $\mathbf{P}^*$ is defined as $\mathbf{P}^*={\rm diag}\{[P_T\,0\,\cdots\,0]\}$, and the optimal UPS ratio is $\rho_{_\mathrm{EB}}$ given by \eqref{eq:reb}. By substituting $\mathbf{S_{_\mathrm{EB}}}$ and $\rho_{_\mathrm{EB}}$ into \eqref{eq:KKT1} and \eqref{eq:KKT2}, the Lagrange multipliers $\mu_{_\mathrm{EB}}$ and $\nu_{_\mathrm{EB}}$ can be written in closed-form as:
\begin{subequations}
\begin{equation}\label{eq:cfmu}
\mu_{_\mathrm{EB}}\triangleq{\sigma^2\,\ln2}\left(1+ \frac{\left(1-\rho_{_\mathrm{EB}}\right)P_T[\boldsymbol{\Lambda}]_{1,1}^2}{\sigma^2}\right), 
\end{equation}
\begin{equation}\label{eq:cfnu}
\nu_{_\mathrm{EB}}\triangleq\frac{\mu_{_\mathrm{EB}}\left(1-\rho_{_\mathrm{EB}}\right)[\boldsymbol{\Lambda}]_{1,1}^2}{\ln(2)\left(\sigma^2+\left(1-\rho_{_\mathrm{EB}}\right)P_T[\boldsymbol{\Lambda}]_{1,1}^2\right)}+[\boldsymbol{\Lambda}]_{1,1}^2 \rho_{_\mathrm{EB}}.
\end{equation}
\end{subequations} 
We therefore conclude that $\left(\mathbf{S^*},\rho^*,\mu^*,\nu^*\right)$ is given as $\left(\mathbf{S}_{_\mathrm{EB}},\rho_{_\mathrm{EB}},\mu_{_\mathrm{EB}},\nu_{_\mathrm{EB}}\right)$ for $R\le{R_{\mathrm{th}}}$. When $R>{R_{\mathrm{th}}}$, the optimum TX precoding operation is spatial multiplexing and we thus apply the following algebraic manipulations to \eqref{eq:KKT1} to obtain the TX covariance matrix:   
\begin{align}\label{eq:KKT1b}
\textstyle \mathbf{H}^{\rm H}\left(\mathbf{I}_{N_R}+\left(1-\rho\right)\sigma^{-2}\mathbf{H}\mathbf{S}\mathbf{H}^{\rm H}\right)^{-1}\mathbf{H}&\stackrel{(a)}{=}{\left(\nu\mathbf{I}_{N_T}-\rho\,\mathbf{H}^{\rm H}\mathbf{H}\right)}{\sigma^2\,\ln2}\left[{\mu\,\left(1-\rho\right)}\right]^{-1},\nonumber\\
\textstyle \left(\mathbf{I}_r+\left(1-\rho\right)\sigma^{-2}\mathbf{\Lambda V}^{\rm H} \mathbf{S\, V\Lambda}\right)^{-1} &\stackrel{(b)}{=}{\left(\nu\,\mathbf{I}_{r}-\rho\,\boldsymbol{\Lambda}^{\rm H}\boldsymbol{\Lambda}\right)\mathbf{\Lambda}^{-2}}{\sigma^2\,\ln2}\left[{\mu\,\left(1-\rho\right)}\right]^{-1},\nonumber\\
\textstyle\mathbf{\Lambda V}^{\rm H} \mathbf{S\, V\Lambda}&\stackrel{(c)}{=}{\frac{\mu}{\ln2}}{\left(\nu\,\mathbf{I}_r-\rho\,\boldsymbol{\Lambda}^{\rm H}\boldsymbol{\Lambda}\right)^{-1}\mathbf{\Lambda}^2}-\frac{\sigma^2}{1-\rho}\,\mathbf{I}_r,
\end{align}
where $(a)$ is obtained after some rearrangements in \eqref{eq:KKT1} and $(b)$ is deduced from the following four operations: \textit{i}) substitution of the reduced SVD of $\mathbf{H}$; \textit{ii}) left multiplication with $\mathbf{V}^{\rm H}$ and right with $\mathbf{V}$; \textit{iii}) left and right multiplication of both sides with $\mathbf{\Lambda}^{-1}$; and \textit{iv}) pushing $\mathbf{U}^{\rm H}$ and $\mathbf{U}$ inside the inverse. Finally, $(c)$ is obtained after taking the inverse of $(b)$ and applying some rearrangements. By performing the necessary left and right multiplications of \eqref{eq:KKT1b} with $\mathbf{\Lambda}^{-1}$, $\mathbf{V}$, and $\mathbf{V}^{\rm H}$ and setting $\rho$, $\mu$, and $\nu$ to their optimal values $\rho_{_{\rm SM}}$, $\mu_{_{\rm SM}}$, and $\nu_{_{\rm SM}}$ for spatial multiplexing, the optimal TX covariance matrix for $R>{R_{\mathrm{th}}}$ can be derived as $\mathbf{S^*}=\mathbf{S_{_\mathrm{SM}}}$, where 
\begin{equation}\label{eq:optS0}
\textstyle\mathbf{S_{_\mathrm{SM}}}=\mathbf{V}\left(\frac{\mu_{_{\rm SM}}}{\ln2}\left(\nu_{_{\rm SM}}\,\mathbf{I}_r-\rho_{_{\rm SM}}\,\boldsymbol{\Lambda}^{\rm H}\boldsymbol{\Lambda}\right)^{-1}-\frac{\sigma^2}{1-\rho_{_{\rm SM}}}\,\mathbf{\Lambda}^{-2}\right)\mathbf{V}^{\rm H}.
\end{equation}

Applying some rearrangements in \eqref{eq:KKT2} to solve for the optimal $\mu_{_{\rm SM}}$ yields 
\begin{align}\label{eq:KKT2b}
\textstyle \mu_{_{\rm SM}} = \frac{\mathrm{tr}\left(\mathbf{H}\mathbf{S_{_\mathrm{SM}}}\mathbf{H}^{\rm H}\right)}{\mathrm{tr}\left(\frac{\mathbf{H}\mathbf{S_{_\mathrm{SM}}}\mathbf{H}^{\rm H}}{\sigma^2\,\ln2} \left(\mathbf{I}_{N_R}+\left(1-\rho_{_{\rm SM}}\right)\sigma^{-2}\mathbf{H}\mathbf{S_{_\mathrm{SM}}}\mathbf{H}^{\rm H}\right)^{-1}\right)}.
\end{align} 
Evidently from \eqref{eq:optS0}, $\mathbf{S_{_\mathrm{SM}}}\in\mathbb{C}^{N_T\times N_T}$ can be expressed as $\mathbf{S_{_\mathrm{SM}}}\triangleq\mathbf{VP^*V}^{\rm H}$ with the $r\times r$ matrix $\mathbf{P^*}\triangleq{\rm diag}\{[p_1^*\, p_2^*\, \cdots\, p_{r}^*]\}$ representing the optimal power allocation matrix among $\mathbf{H}$'s eigenchannels. So, the optimal power assignment of the $k$-th eigenchannel is given by
\begin{align}\label{eq:PA}
\textstyle p_k^*=\left(\frac{\mu^*}{\ln2\left(\nu_{_{\rm SM}}-\rho_{_{\rm SM}}[\boldsymbol{\Lambda}]_{k,k}^2\right)} -\frac{\sigma^2}{\left(1-\rho_{_{\rm SM}}\right)[\boldsymbol{\Lambda}]_{k,k}^2}\right)^+,\quad\forall\,k=1,2,\ldots,r.
\end{align} 
The optimal $\rho_{_{\rm SM}}$, $\mu_{_{\rm SM}}$, and $\nu_{_{\rm SM}}$ is the solution of the system with the three equations \eqref{eq:KKT3}, $\eqref{eq:KKT4}$, and \eqref{eq:KKT2b} after setting $\mathbf{S}=\mathbf{S_{_\mathrm{SM}}}$ and satisfying $\mu,\nu>0$ and $0\le\rho<1$. {Later in Section~\ref{sec:EGoA} we first reduce this system of equations to two, and then by exploiting the tight bounds on $\nu^*$ derived in Section~\ref{sec:Multipliers_bounds}, we present how it can be implemented as an efficient 2-D linear search.}
\begin{remark}\label{rem:eig}
Observing \eqref{eq:optS0} and \eqref{eq:PA} leads to the conclusion that the optimum TX precoding for ${R}>{R_{\mathrm{th}}}$ is $\mathbf{F}=\mathbf{V}({\mathbf{P}^*})^{1/2}$ with the $r$ diagonal elements of ${\mathbf{P}^*}$ given by \eqref{eq:PA}. This precoding results in $r$ parallel eigenchannel transmissions with power allocation obtained from a modified waterfilling algorithm, where the different water levels depend on ${R}$, $P_T$, $\mathbf{H}$, and $\sigma^2$.   
\end{remark}  
By combining the optimal TX covariance matrices for both cases of energy beamforming and spatial multiplexing, the globally optimal solution $\left(\mathbf{S^*},\rho^*\right)$ for $\mathcal{OP}1$ can be summarized as
\begin{eqnarray}\label{eq:optS}
\mathbf{S^*} = \begin{cases}
P_T\,\mathbf{v}_1\mathbf{v}_1^{\rm H},   & \text{${R}\le{R_{\mathrm{th}}}\le R_{\max}$},\\
\mathbf{V}\left(\frac{\mu^*}{\ln2}\,\left(\nu^*\mathbf{I}_r-\rho\,\boldsymbol{\Lambda}^{\rm H}\boldsymbol{\Lambda}\right)^{-1}-\frac{\sigma^2}{1-\rho}\,\mathbf{\Lambda}^{-2}\right)\mathbf{V}^{\rm H},   & \text{${R_{\mathrm{th}}}<R\le R_{\max}$},\\
\text{Infeasible}, & \text{${R}>{R_{\max}}$},
\end{cases}
\end{eqnarray}
where $\rho^*=\rho_{_\mathrm{EB}}$, $\mu^*=\mu_{_\mathrm{EB}}$, and $\nu^*=\nu_{_\mathrm{EB}}$ for ${R}\le{R_{\mathrm{th}}}$, and for ${R}>{R_{\mathrm{th}}}$, $\rho^*=\rho_{_\mathrm{SM}}$, $\mu^*=\mu_{_{\rm SM}}$, and $\nu^*=\nu_{_{\rm SM}}$ are obtained from the solution of the system of equations described below \eqref{eq:PA}. \color{black} The feasibility of $\mathcal{OP}1$ depends on ${R_{\max}}\triangleq\log_2\left(\det\left(\mathbf{I}_{N_R}+\sigma^{-2}\mathbf{H}\mathbf{S_{_\mathrm{WF}}}\mathbf{H}^{\rm H}\right)\right)$, which represents the maximum achievable rate for UPS ratio $\rho=0$ and $\mathbf{S_{_\mathrm{WF}}}\triangleq \mathbf{V}\mathbf{P_{_\mathrm{WF}}}\mathbf{V}^{\rm H}$. In the latter expression, $\mathbf{P_{_\mathrm{WF}}}\triangleq{\rm diag}\{[p_{_{\mathrm{WF},1}}\, p_{_{\mathrm{WF},2}}\, \cdots\, p_{_{\mathrm{WF},r}}]\}$ is the $r\times r$ power allocation matrix whose rank $r_w$ (non-zero diagonal entries) is given by~\cite{WFG}
\begin{eqnarray}\label{eq:rnkWF}
\textstyle r_w\triangleq \max\left\lbrace k\,\mathrel{}\middle|\mathrel{}\left(P_T-\sum\limits_{i=1}^{k-1}\left(\frac{\sigma^2}{[\boldsymbol{\Lambda}]_{k,k}^2}-\frac{\sigma^2}{[\boldsymbol{\Lambda}]_{i,i}^2}\right)\right)^{\hspace{-1.5mm}+}\hspace{-2mm}>0,\, 1\le k\le r\hspace{-1mm}\right\rbrace,
\end{eqnarray} 
and its non-zero elements are obtained from the standard waterfilling algorithm as
\begin{eqnarray}\label{eq:PAWF}
{p_{_{\mathrm{WF},k}}=\begin{cases}
r_w^{-1}\left({P_T-\sum\limits_{i=1}^{r_w-1}\left(\frac{\sigma^2}{[\boldsymbol{\Lambda}]_{r_w,r_w}^2}-\frac{\sigma^2}{[\boldsymbol{\Lambda}]_{i,i}^2}\right)}\right)+\frac{\sigma^2}{[\boldsymbol{\Lambda}]_{r_w,r_w}^2}-\frac{\sigma^2}{ [\boldsymbol{\Lambda}]_{k,k}^2}, & \text{$k=1,2,\ldots,r_w$}\\
0, & \text{$r_w+1\le k\le r$}
\end{cases}}.
\end{eqnarray}  
 
{Here we would like to add that based on \eqref{eq:optS} deciding whether the optimal TX precoding matrix $\mathbf{S}^*$ is denoted $\mathbf{S_{_\mathrm{EB}}}$ or $\mathbf{S_{_\mathrm{SM}}}$, the corresponding optimal TX signal vector $\mathbf{x}^*\in\mathbb{C}^{N_T\times 1}$ can be obtained as $\mathbf{x_{_\mathrm{EB}}}\triangleq\sqrt{P_T}\,\mathbf{v}_1\,\widetilde{x}$ and $\mathbf{x_{_\mathrm{SM}}}\triangleq\mathbf{V}\left({\mathbf{P}^*}\right)^{1/2}\,\widetilde{\mathbf{x}}$. Here  $\widetilde{x}$ is an arbitrary ZMCSCG random signal and $\widetilde{\mathbf{x}}\in\mathbb{C}^{r\times 1}$ is a ZMCSCG random  vector, both having unit variance entries.}

\subsection{Globally Optimal Solution of $\mathcal{OP}2$}\label{sec:gop2}
{Like} $\mathcal{OP}1$, there exists a rate threshold {in} $\mathcal{OP}2$ that determines whether energy beamforming or spatial multiplexing is the optimal TX precoding operation. This value is given by 
\begin{equation}\label{eq:Rid}
{R_{\mathrm{th}}^{\mathrm{id}}}\triangleq\log_2\left(1+{\sigma^{-2}}{P_T\,[\boldsymbol{\Lambda}]_{1,1}^2}\right),
\end{equation}
which represents the rate achieved by energy beamforming in the ideal reception case.
\begin{lemma}\label{lem:GOP2}
The globally optimal solution $\mathbf{S_{\text{id}}^*}$ of $\mathcal{OP}2$ is given by
\begin{eqnarray}\label{eq:optS2b3}
\hspace{-2mm}\mathbf{S_{\text{id}}^*}= \begin{cases}
P_T\,\mathbf{v}_1\mathbf{v}_1^{\rm H},   & \text{${R}\le{R_{\mathrm{th}}^{\mathrm{id}}}\le R_{\max}$,}\\
\mathbf{V}{\rm diag}\{[p_1^{(\rm id)}\,p_2^{(\rm id)}\,\cdots\,p_r^{(\rm id)}]\}\mathbf{V}^{\rm H},   & \text{${R_{\mathrm{th}}^{\mathrm{id}}}<{R}\le R_{\max}$,}\\
\text{Infeasible}, & \text{${R}>{R_{\max}}$},
\end{cases}
\end{eqnarray} 
where {$p_k^{(\rm id)}$ denotes the power assignment of the $k$-th eigenchannel and is given by
	
\noindent\begin{equation}\label{eq:TX_optS2a}
p_k^{(\rm id)} = \left(\frac{\mu_2^*}{\ln2\left(\nu_2^*-[\boldsymbol{\Lambda}]_{k,k}^2\right)} -\frac{\sigma^2}{[\boldsymbol{\Lambda}]_{k,k}^2}\right)^+,\quad \forall\, k=1,2,\ldots,r.
\end{equation}}
In the latter expression, $\nu_2^*\ge0$ and $\mu_2^*\ge0$ represent the Lagrange multipliers corresponding to constraints $({\rm C2})$ and $({\rm C5})$, respectively. These can be obtained using a subgradient method as described in \cite[App. A]{MIMO_SWIPT} such that $\log_2\left(\det\left(\mathbf{I}_{N_R}+\sigma^{-2}\mathbf{H}\mathbf{S}_{\text{id}}^*\mathbf{H}^{\rm H}\right)\right)={R}$ and $\mathrm{tr}\left(\mathbf{S}_{\text{id}}^*\right)= P_T$. 
\end{lemma} 
\begin{IEEEproof}
The proof is provided in Appendix~\ref{App:GOP2}.
\end{IEEEproof}
\color{black}
\begin{remark}{It can be observed from the solution of $\mathcal{OP}2$ that our proposed TX precoding design significantly differs from that obtained from the solution of  optimization problem $({\rm P}3)$ in \cite{MIMO_SWIPT}. This reveals that the design maximizing the total received RF power for EH, while satisfying a minimum instantaneous rate requirement, is very different from the design that maximizes the instantaneous rate subject to a minimum constraint on the total received RF power.} 
\end{remark}\color{black}

\section{Analytical Bounds and Asymptotic Approximations}\label{sec:ana}
Here we first present analytical bounds for the UPS ratio $\rho$ and the Lagrange multipliers $\mu$ and $\nu$ appearing in the KKT conditions for both optimization problems considered in Section~\ref{sec:soln}. Then tight approximations for high SNR values for the optimal TX precoding designs  are presented.

\subsection{Analytical Bounds}
\subsubsection{UPS Ratio $\rho$}\label{sec:UPS_Ratio_bounds}
The information rate is given by $\log_2\left(\det\left(\mathbf{I}_{N_R}+\left(1-\rho\right)\sigma^{-2}\mathbf{H}\mathbf{S}\mathbf{H}^{\rm H}\right)\right)$, which is a monotonically decreasing function of $\rho$. The upper bound $\rho_{_\mathrm{UB}}$ on the feasible $\rho$ value satisfying the rate constraint $({\rm C1})$ is given by the UPS ratio corresponding to the maximum achievable rate value $R_{\max}$. This maximum value is achieved with statistical multiplexing over all available eigenchannels. In mathematical terms, $\rho_{_\mathrm{UB}}$ can be obtained by setting $\mathbf{S}$ as $\mathbf{S_{_\mathrm{WF}}}=\mathbf{V}\mathbf{P_{_\mathrm{WF}}}\mathbf{V}^{\rm H}$ with the entries of $\mathbf{P_{_\mathrm{WF}}}$ defined in \eqref{eq:PAWF} yielding
\begin{align}\label{eq:rho-UB}
\rho_{_\mathrm{UB}}\triangleq & \,\bigg\{ \rho \;\bigg| \det\left(\mathbf{I}_{N_R}+\sigma^{-2}{\left(1-\rho\right)\mathbf{H}\mathbf{S_{_\mathrm{WF}}}\mathbf{H}^{\rm H}}\right)=2^{{R}} \,\,{\rm and}\,\, 0\le\rho\le 1\bigg\}. 
\end{align} 
Likewise, the lower bound on the feasible $\rho$ value meeting constraint $({\rm C1})$ is given by the UPS ratio $\rho_{_\mathrm{EB}}$ as defined in \eqref{eq:reb}. This lower bound happens with energy beamforming, where the entire TX power is allocated to the best gain eigenchannel and the achievable rate is minimum. Combining the latter derivation results in $\rho_{_\mathrm{EB}}\le\rho\le\rho_{_{\mathrm{UB}}}$.

\subsubsection{Lagrange Multipliers $\mu$ and $\nu$ for ${R}>{R_{\mathrm{th}}}$}\label{sec:Multipliers_bounds}
To have non-negative power allocation $p_1$ over the best gain eigenchannel having eigenmode $[\boldsymbol{\Lambda}]_{1,1}$, it must hold from \eqref{eq:PA} that $\nu\ge\nu_{_\mathrm{LB}}\triangleq\rho\,[\boldsymbol{\Lambda}]_{1,1}^2$. Also, using the definition $p_1=\alpha P_T$ with $\alpha\le1$ in \eqref{eq:PA} for $k=1$ yields ${\mu} = \left(\nu-\rho[\boldsymbol{\Lambda}]_{1,1}^2\right){\left(\alpha\,P_T+\frac{\sigma^2}{\left(1-\rho\right)[\boldsymbol{\Lambda}]_{1,1}^2}\right)\ln2}$. Since for the total received power holds $\mathrm{tr}\left(\mathbf{HSH}^{\rm H}\right)\le$

\noindent$P_T$ and also \eqref{eq:KKT2b} holds, the upper bound for $\mu$, denoted by $\mu_{_\mathrm{UB}}$, can be obtained as
\begin{align} \label{eq:muub}
\mu&=\frac{\mathrm{tr}\left(\mathbf{HSH}^{\rm H}\right)}{\mathrm{tr}\left(\frac{\mathbf{H}\mathbf{S}\mathbf{H}^{\rm H}}{\sigma^2\,\ln2} \left(\mathbf{I}_{N_R}+\left(1-\rho\right)\sigma^{-2}\mathbf{H}\mathbf{S}\mathbf{H}^{\rm H}\right)^{-1}\right)} {\stackrel{(d)}{<}}\,\frac{\mathrm{tr}\left(\mathbf{HSH}^{\rm H}\right)}{\frac{r}{\left(1-\rho\right)\ln2}}<\mu_{_\mathrm{UB}}\triangleq\frac{\left(1-\rho\right)P_T\ln2}{r},
\end{align}
where $(d)$ results from the high SNR approximation. Combining   \eqref{eq:muub} with  $\frac{\sigma^2}{\left(1-\rho\right)[\boldsymbol{\Lambda}]_{1,1}^2}> 0$,  leads to $\nu< \frac{\left(1-\rho\right)}{\alpha\,r}+\rho[\boldsymbol{\Lambda}]_{1,1}^2$. Due to the highest power allocation over the best gain eigenchannel, it must hold $\alpha\ge \frac{1}{r}$, yielding $\nu_{_\mathrm{UB}}\triangleq 1+\rho([\boldsymbol{\Lambda}]_{1,1}^2-1)$. {However as shown later, $\nu\ll\nu_{_\mathrm{UB}}$ because the total received power $P_R$ is usually much less than $P_T$. These analytical bounds will be used in Section~\ref{sec:2D} for presenting an efficient implementation of the global optimization algorithm.}

\subsection{Asymptotic Analysis} 

As discussed in~\cite{Niyato_Survey1, ComMag} the received RF power for EH in SWIPT systems needs to be greater than energy reception sensitivity, which is in the order of $-10$dBm to $-30$dBm, for the practical RF EH circuits to provide non-zero harvested DC power after rectification. Since the received noise power spectral density is around $-175$dBm/Hz leading to an average received noise power of around $-100$dBm for SWIPT at $915$ MHz, the received SNR in practical SWIPT systems is very high, i$.$e$.$, around $70$dB, even for very high frequency transmissions. Based on this practical observation for SWIPT systems, we next investigate the joint design for high SNR scenarios.

\subsubsection{Globally Optimal Solution of $\mathcal{OP}1$ for High SNR}
The globally optimal solution of $\mathcal{OP}1$ for high SNR values defined as $\left(\mathbf{S_a^*},\rho_a^*,\mu_a^*,\nu_a^*\right)$ can be obtained similarly to Section~\ref{sec:KKT} as 
\begin{eqnarray}\label{eq:S_OP1_highSNR}
\mathbf{S_{\mathrm{a}}^*} = \begin{cases}
P_T\,\mathbf{v}_1\mathbf{v}_1^{\rm H},   & \text{${R}\le{R_{\mathrm{th}}}\le R_{\max},$ }\\
\mathbf{V}\left(\frac{\mu_a^*}{\ln2}\,\left(\nu_a^*\mathbf{I}_r-\rho_a^*\,\boldsymbol{\Lambda}^{\rm H}\boldsymbol{\Lambda}\right)^{-1}\right)\mathbf{V}^{\rm H},   & \text{${R_{\mathrm{th}}}<R\le R_{\max}$ },\\
\text{Infeasible}, & \text{${R}>{R_{\max}}$},
\end{cases} 
\end{eqnarray}  
$
\mu_a^* = r^{-1}\left(1-\rho_a^*\right)\mathrm{tr}\left(\mathbf{HS_a^*H}^{\rm H}\right)\ln2,$
and the remaining two unknowns $\rho_a^*$ and $\nu_a^*$ are given from the solutions of the equations $\mathrm{tr}\left(\mathbf{S_a^*}\right)=P_T$ and  $\log_2\left(\det\left(\sigma^{-2}\left(1-\rho\right)\mathbf{HS_a^*H}^{\rm H}\right)\right)={R}$. After some simplifications with \eqref{eq:S_OP1_highSNR}, the power allocation is obtained as $p_{a,k}^*= \frac{\mu_a^*}{\left(\nu_a^*-\rho_a^*[\boldsymbol{\Lambda}]_{k,k}^2\right)\ln2}, \forall k=1,2,\ldots,r.$
Hence, under high SNR, the optimal power allocation over available eigenchannels for ${R}>{R_{\mathrm{th}}}$ is always greater than zero regardless of the relative strengths of the eigenmodes.

\subsubsection{Globally Optimal Solution of $\mathcal{OP}2$ for High SNR}\label{sec:cf-asym}
By using the previously derived analytical bounds for $\rho$ and $\nu$ along with Lemma~\ref{lem:GOP2}, the approximation $\mathbf{{S_{\text{id,a}}^*}}$ for the globally optimal TX covariance matrix $\mathbf{{S_{\text{id}}^*}}$ of $\mathcal{OP}2$ for high SNR values can be obtained as
\begin{eqnarray}
\hspace{-2mm}\mathbf{{S_{\text{id,a}}^*}}= \begin{cases}
P_T\,\mathbf{v}_1\mathbf{v}_1^{\rm H},   & \text{${R}\le{R_{\mathrm{th}}^{\mathrm{id}}}\le R_{\max}$,}\\
\mathbf{V}{\rm diag}\{[p_1^{(\rm id,a)}\,p_2^{(\rm id,a)}\,\cdots\,p_r^{(\rm id,a)}]\}\mathbf{V}^{\rm H},   & \text{${R_{\mathrm{th}}^{\mathrm{id}}}<{R}\le R_{\max}$,}\\
\text{Infeasible}, & \text{${R}>{R_{\max}}$},
\end{cases}
\end{eqnarray} 
where each $p_k^{(\rm id,a)}$ with $k=1,2,\ldots,r$ is given by
\begin{equation}\label{eq:PA-a-I}
p_k^{(\rm id,a)} = \frac{\mu_{2a}^*}{\left(\nu_{2a}^*-[\boldsymbol{\Lambda}]_{k,k}^2\right)\ln2}.
\end{equation}
With $[\boldsymbol{\Lambda}]_{1,1}^2<\nu_{2a}^*<[\boldsymbol{\Lambda}]_{1,1}^2+ 1$, solving $\displaystyle\sum_{k=1}^{r}p_k^{(\rm id,a)}=P_T$ yields $
\mu_{2a}^*  = P_T\left(\textstyle\sum\limits_{k=1}^{r} \frac{1}{\left(\nu_{2a}^*-[\boldsymbol{\Lambda}]_{k,k}^2\right)\ln2}\right)^{-1}.$ 
We now set $\beta\triangleq\nu_{2a}^*-[\boldsymbol{\Lambda}]_{1,1}^2\in(0,1)$ and substitute into \eqref{eq:PA-a-I} in order to rewrite each $p_k^{(\rm id,a)}$ as
\begin{align} \label{eq:PA-a-I20}
p_k^{(\rm id,a)} = \frac{\beta \, p_1^{(\rm id,a)}}{\beta+[\boldsymbol{\Lambda}]_{1,1}^2-[\boldsymbol{\Lambda}]_{k,k}^2}. 
\end{align}

To solve for $\beta$, we need to replace into the rate constraint expression leading to $\prod\limits_{k=1}^{r}\left(\frac{p_k^{(\rm id,a)}\,[\boldsymbol{\Lambda}]_{k,k}^2}{\sigma^2}\right)$ $=2^{{R}}$, which after some mathematical simplifications results in the expression
\begin{align}\label{eq:PA-a-I2}
\textstyle\prod\limits_{k=1}^{r} \left(\frac{\beta\,[\boldsymbol{\Lambda}]_{k,k}^2}{\beta+[\boldsymbol{\Lambda}]_{1,1}^2-[\boldsymbol{\Lambda}]_{k,k}^2}\right)=2^{{R}}\left(\frac{\sigma^2}{p_1^{(\rm id,a)}}\right)^{r}.
\end{align}
The $p_1^{(\rm id,a)}$ included in \eqref{eq:PA-a-I2} can be obtained in closed-form as a function of $\beta$ by solving $\textstyle\sum_{k=1}^{r}p_k^{(\rm id,a)}=P_T$ and using \eqref{eq:PA-a-I20}, yielding
\begin{align} \label{eq:PA-a-I3}
p_1^{(\rm id,a)}= {P_T}\left({1+\textstyle\sum\limits_{k=2}^{r}\left(\frac{\beta}{\beta+[\boldsymbol{\Lambda}]_{1,1}^2-[\boldsymbol{\Lambda}]_{k,k}^2}\right)}\right)^{-1}.
\end{align}
{Using these developments in Section~\ref{sec:asymimp} we show that the asymptotically optimal TX precoding for $\mathcal{OP}2$ can be obtained using a 1-D linear search over very short range $\left(0,1\right)$ of $\beta$.}

\begin{remark}\label{rem:asym}
With the expressions \eqref{eq:PA-a-I2} and \eqref{eq:PA-a-I3} resulted from our derived asymptotic analysis, we have managed to replace the problem of finding the positive real values of the Lagrange multipliers $\mu_2$ and $\nu_2$ in $\mathcal{OP}2$ along with the required waterfilling-based decision making process \big(this process involves the discontinuous function $\left(x\right)^+$ due to the implicit consideration of constraint $({\rm C3})$\big) by a simple linear search for parameter $\beta$ belonging in the range $\left(0,1\right)$.
\end{remark}

\section{Efficient Global Optimization Algorithm}\label{sec:EGoA}
The goal of this section is to first present a global optimization algorithm to obtain the previously derived globally optimal solutions for $\mathcal{OP}1$ and $\mathcal{OP}2$ by effectively solving the KKT conditions. After that we present an alternate low complexity algorithm based on a simple 2-D linear search to practically implement the former algorithm in a computational efficient and analytically tractable manner while meeting a desired level of accuracy.
  
\subsection{Solving the KKT Conditions}
As discussed in Section~\ref{sec:soln}, the globally optimal $\mathbf{S^*}$ and $\rho^*$ for ${R}>{R_{\mathrm{th}}}$ of $\mathcal{OP}1$ are obtained by solving the system of three equations \eqref{eq:KKT3}, \eqref{eq:KKT4}, and \eqref{eq:KKT2b} for $\rho^*,\mu^*,$ and $\nu^*$ after setting $\mathbf{S}=\mathbf{S_{_\mathrm{SM}}}$. Likewise, as presented in Lemma~\ref{lem:GOP2}, the globally optimal $\mathbf{S_{\text{id}}^*}$ for ${R}>{R_{\mathrm{th}}^{\mathrm{id}}}$ of $\mathcal{OP}2$ is derived by solving $\log_2\left(\det\left(\mathbf{I}_{N_R}+\sigma^{-2}\mathbf{H}\mathbf{S}_{\text{id}}^*\mathbf{H}^{\rm H}\right)\right)={R}$ and $\mathrm{tr}\left(\mathbf{S}_{\text{id}}^*\right)= P_T$ for $\mu_2^*$ and $\nu_2^*$.

\subsubsection{Reduction of the System of Non-linear Equations}\label{sec:red1}
It is in general very difficult to efficiently solve a large system of non-linear equations. Hereinafter, we discuss the reduction of the number of the non-linear equations to be solved from three to two in $\mathcal{OP}1$ and from two to one in $\mathcal{OP}2$.

Let us denote the rank of the optimal TX covariance matrix by $r_s$. {It} represents the number of eigenchannels that have non-zero power allocation, i.e., $p_k>0$ with $k=1,2,\ldots,r_s$. Substituting this definition into \eqref{eq:KKT4} and \eqref{eq:PA} with $\nu>0$, we can express $\mu^*$ in terms of $\nu^*$ and $\rho^*$ as
\begin{align}\label{eq:mu1a}
 \mu^*=\frac{P_T+\textstyle\sum_{k=1}^{r_s} {\sigma^2}\left(\left(1-\rho^*\right)[\boldsymbol{\Lambda}]_{k,k}^2\right)^{-1}}{r_s\textstyle\sum_{k=1}^{r_s}\left(\left(\nu^*-\rho^*[\boldsymbol{\Lambda}]_{k,k}^2\right)\ln2\right)^{-1}}.
\end{align}  
{Using} the definition of $r_s$ in \eqref{eq:KKT3} and \eqref{eq:PA} with $\mu>0$, $\mu^*$ can be alternatively expressed as
\begin{align}\label{eq:mu1b}
 \mu^*=2^{\frac{{R}}{r_s}}{\sigma^2}\left(\left(1-\rho^*\right)\left(\textstyle\prod_{k=1}^{r_s}\frac{[\boldsymbol{\Lambda}]_{k,k}^2}{\nu^*-\rho^*[\boldsymbol{\Lambda}]_{k,k}^2}\right)^{\frac{1}{r_s}}\right)^{-1}\ln2.
\end{align}  
By combining \eqref{eq:KKT2b}, \eqref{eq:mu1a}, and \eqref{eq:mu1b}, the reduced system of two non-linear equations to be solved for $\rho^*$ and $\nu^*$ as included in the KKT point $\left(\mathbf{S^*},\rho^*,\mu^*,\nu^*\right)$ for ${R}>{R_{\mathrm{th}}}$ in $\mathcal{OP}1$ is given by
\begin{subequations}
\begin{equation}\label{eq:s1}
\frac{P_T\left(1-\rho^*\right)\sigma^{-2}+\sum_{k=1}^{r_s} [\boldsymbol{\Lambda}]_{k,k}^{-2}}{r_s\sum_{k=1}^{r_s}\left(\nu^*-\rho^*[\boldsymbol{\Lambda}]_{k,k}^2\right)^{-1}}=2^{\frac{{R}}{r_s}}\left(\textstyle\prod\limits_{k=1}^{r_s}\frac{[\boldsymbol{\Lambda}]_{k,k}^2}{\nu^*-\rho^*[\boldsymbol{\Lambda}]_{k,k}^2}\right)^{-\frac{1}{r_s}},
\end{equation}
\begin{equation}\label{eq:s2}
2^{\frac{{R}}{r_s}}\sigma^2\left(\textstyle\sum\limits_{k=1}^{r_s}\frac{p_k^*\,[\boldsymbol{\Lambda}]_{k,k}^2}{1+\left(1-\rho^*\right)p_k^*\,[\boldsymbol{\Lambda}]_{k,k}^2}\right)\ln2=\left(1-\rho^*\right){\textstyle\sum\limits_{k=1}^{r_s}p_k^*\,[\boldsymbol{\Lambda}]_{k,k}^2}{\left(\textstyle\prod\limits_{j=1}^{r_s}\frac{[\boldsymbol{\Lambda}]_{j,j}^2}{\nu^*-\rho^*[\boldsymbol{\Lambda}]_{j,j}^2}\right)^{\frac{1}{r_s}}},
\end{equation}
\end{subequations}
where $p_k^*=\frac{\sigma^2}{\left(1-\rho^*\right)}\left({{{2^{\frac{{R}}{r_s}}}}\left(\left(\nu^*-\rho^*[\boldsymbol{\Lambda}]_{k,k}^2\right){{\left(\prod_{j=1}^{r_s}\frac{[\boldsymbol{\Lambda}]_{j,j}^2}{\nu^*-\rho^*[\boldsymbol{\Lambda}]_{j,j}^2}\right)^{\frac{1}{r_s}}}}\right)^{-1}} -\frac{1}{[\boldsymbol{\Lambda}]_{k,k}^2}\right)^+$ $\forall$$k=1,2,\ldots,r_s$. 

In a similar manner, the single non-linear equation that needs to be solved for computing $\nu_2^*$ included in the KKT point $\left(\mathbf{S_{\text{id}}^*},\mu_2^*,\nu_2^*\right)$ for ${R}>{R_{\mathrm{th}}^{\mathrm{id}}}$ in $\mathcal{OP}2$ is given by

\noindent\begin{equation}\label{eq:s1-I}
{\left(\textstyle\frac{P_T}{\sigma^2}+\textstyle\sum\limits_{j=1}^{r_s}\frac{1}{ [\boldsymbol{\Lambda}]_{j,j}^2}\right)}{\displaystyle\left(\textstyle\prod\limits_{k=1}^{r_s}\frac{[\boldsymbol{\Lambda}]_{k,k}^2}{\nu_2^*- [\boldsymbol{\Lambda}]_{k,k}^2}\right)^{\frac{1}{r_s}}}={\textstyle2^{\frac{{R}}{r_s}}}\,{r_s \textstyle\sum\limits_{k=1}^{r_s} \left(\nu_2^*-[\boldsymbol{\Lambda}]_{k,k}^2\right)^{-1}}.
\end{equation}

\begin{lemma}\label{lem:rnk}
The rank $r_s$ of the optimal TX covariance matrix $\mathbf{S^*}$ of $\mathcal{OP}1$ (or $\mathbf{S_{\text{id}}^*}$ of $\mathcal{OP}2$) is always lower or equal to the rank $r_w$ of  $\mathbf{S_{_\mathrm{WF}}}$ providing the maximum achievable rate ${R_{\max}}$.
\end{lemma}
\begin{IEEEproof}
The proof follows from the discussion in Section~\ref{sec:KKT}. The maximum received power $P_{R,E}$ for EH is given by the rank-$1$ covariance matrix $\mathbf{S_{_\mathrm{EB}}}$ implying TX energy beamforming. With increasing rate requirement ${R}>{R_{\mathrm{th}}}$, the optimal TX precoding switches from energy beamforming to statistical multiplexing. In this case, the power allocated over the best gain eigenchannel monotonically decreases due to the power allocation among the other available eigenchannels, thus increasing $r_s$ and decreasing $P_{R,E}$. The rate $R_{\max}$ is achieved by $\mathbf{S_{_\mathrm{WF}}}$ having rank $r_w\le r$, which also results in the minimum $P_{R,E}$ for both $\mathcal{OP}1$ and $\mathcal{OP}2$. Therefore, $r_w$ represents the maximum rank of the TX covariance matrix, hence it must hold $r_s\le r_w$. 
\end{IEEEproof}

\subsubsection{Implementation Details and Challenges}\label{sec:impK}
{Here we first present the detailed steps involved in the implementation of solving the reduced system of non-linear equations to obtain the optimal design for both $\mathcal{OP}1$ and $\mathcal{OP}2$ via Algorithm~\ref{Algo:1}. After that we discuss the practical challenges involved in implementing it directly using the commercially available numerical solvers which may suffer from slow convergence issues as faced by the subgradient methods~\cite{MIMO_SWIPT,DPS,Spatial,Close} and semidefinite relaxations~\cite{MISO-PS,SOCP,MUTxMIMO,ASUPS,FDPowerMn}  used in the existing MIMO SWIPT literature.} 

From Algorithm~\ref{Algo:1}, we note that obtaining $\mathbf{S^*}$ and $\rho^*$ involves solving the two non-linear equations \eqref{eq:s1} and \eqref{eq:s2} for at most $r$ times, while considering positive power allocation over the $k$ best gain eigenchannels with $k=1,2,\ldots,r$. Since constraints $({\rm C3})$ and $({\rm C4})$ had been kept implicit, we repeatedly solve the latter system of equations for at most $r_w\le r$ times till we obtain a feasible non-negative power allocation, i$.$e$.$, $\mathbf{S^*}\succeq0$ and $0\le\rho^*\le1$.  

\begin{algorithm}[!h]
\small{\caption{\small Efficient Solution of the KKT Conditions}\label{Algo:1}
	\begin{algorithmic}[1]
		\Require $\mathbf{H}$, $\sigma^2$, $P_T$, and ${R}$ 
		\Ensure Maximized received power $P_{R,E}^*$ for EH along with optimal $\mathbf{S^*}$ and $\rho^*$
		\Statex \textbf{(A) Initialization}
		\State Obtain SVD, $\mathbf{H} = \mathbf{U}\,\mathrm{diag}\{[\boldsymbol{\Lambda}]_{1,1}\,[\boldsymbol{\Lambda}]_{2,2}\,\cdots\,[\boldsymbol{\Lambda}]_{r,r}]\}\,\mathbf{V}^{\rm H}$, along with $r_w$ and $\{p_{_{\mathrm{WF},k}}\}_{k=1}^r$  using \eqref{eq:rnkWF} and \eqref{eq:PAWF} \label{step:in1}
		\State \parbox[t]{\dimexpr\linewidth\relax}{Set ${R_{\max}}=\log_2\left(\det\left(\mathbf{I}_{N_R}+\sigma^{-2}{\mathbf{H}\mathbf{V}{\rm diag}\{[p_{_{\mathrm{WF},1}}\,p_{_{\mathrm{WF},2}}\,\ldots\,p_{_{\mathrm{WF},r}}]\}\mathbf{V}^{\rm H}\mathbf{H}^{\rm H}}\right)\right)$\strut}\label{step:in2}
		\State Set $p_1^*=P_T-P_{\delta}$ and $p_2^*=P_{\delta}$ with $P_{\delta} =10^{-3}$, and obtain ${R_{\mathrm{th}}}$ using \eqref{eq:rth}  \label{step:in3}
		\Statex \textbf{(B) Main Body}
		\If{${R}>{R_{\max}}$}\Comment{Infeasible case}
		\State \Return with an error message that $\mathcal{OP}1$ (or $\mathcal{OP}2$) is infeasible\label{step:in6}
		\ElsIf{${R}\le{R_{\mathrm{th}}}\le{R_{\max}}$}\Comment{Energy Beamforming mode}
		\State Set $\mathbf{S^*}=P_T\,\mathbf{v}_1\mathbf{v}_1^{\rm H}$, $\rho^*=\rho_{_\mathrm{EB}}$, $\mu^*=  \mu_{_\mathrm{EB}}$, and $\nu^*= \nu_{_\mathrm{EB}}$ \label{step:I1}
		\Else\Comment{Modified Statistical Multiplexing mode}
		\State Set $r_s=r_w+1$ 
		\Repeat \mbox{  (Recursion)}
		\State Set $r_s=r_s-1$ 
		\If{$r_s=1$}\State Set $\mathbf{S^*}=P_T\,\mathbf{v}_1\mathbf{v}_1^{\rm H}$,  $\rho^*=\rho_{_\mathrm{EB}}$,  $\mu^*=  \mu_{_\mathrm{EB}}$, and $\nu^*= \nu_{_\mathrm{EB}}$\label{step:I1b}
		\Else
		\State \parbox[t]{\dimexpr\linewidth-\algorithmicindent-\algorithmicindent-\algorithmicindent\relax}{Solve the system of two equations given by  \eqref{eq:s1} and \eqref{eq:s2} to obtain $\rho^*$ and $\nu^*$\strut}\label{step:I2}
		\State Obtain $\mu^*$ by substituting $\rho^*$ and $\nu^*$ in \eqref{eq:mu1a}\label{step:I3}
		\State \parbox[t]{\dimexpr\linewidth-\algorithmicindent-\algorithmicindent-\algorithmicindent\relax}{Set $\mathbf{S^*}=\mathbf{V}{\rm diag}\{[p_1^*\,p_2^*\,\ldots\,p_{r}^*]\}\mathbf{V}^{\rm H}$ with $p_k^*=\begin{cases}
			\frac{\mu^*}{\ln2\left(\nu^*-\rho^*[\boldsymbol{\Lambda}]_{k,k}^2\right)} -\frac{\sigma^2}{\left(1-\rho^*\right)[\boldsymbol{\Lambda}]_{k,k}^2}, & \text{$k=1,2,\ldots,r_s$}\\
			0, & \text{$r_s+1\le k\le r$}.
			\end{cases}$\strut}\label{step:I4}
		\EndIf
		\Until{$\left(p_k^*<0\,\,\forall k=1,2,\ldots,r_s\right)$}
		\EndIf 
	\end{algorithmic}} 
\end{algorithm}	 

Algorithm~\ref{Algo:1} can be slightly modified to provide the globally optimal solution of $\mathcal{OP}2$. In particular, steps~\ref{step:I1},~\ref{step:I1b},~\ref{step:I2},~\ref{step:I3}, and~\ref{step:I4} need to be updated for $\mathcal{OP}2$. Starting with steps~\ref{step:I1} and~\ref{step:I1b}, we need to remove $\rho^*$ since $\mathcal{OP}2$ involves ideal reception and the optimal values of Lagrange multipliers $\mu_2$ and $\nu_2$ for ${R}\le{R_{\mathrm{th}}}$ are given by $\mu_2^*=0$ and $\nu_2^*=[\boldsymbol{\Lambda}]_{1,1}^2$. In addition, to find $\nu_2^*$ for ${R}>{R_{\mathrm{th}}}$ in step~\ref{step:I2} we need to solve \eqref{eq:s1-I}. The solution $\nu_2^*$ of \eqref{eq:s1-I} needs then to update steps~\ref{step:I3} and~\ref{step:I4} in Algorithm~\ref{Algo:1}, and the optimal $\mu_2^*$ and $p_k^{(\rm id)}$'s can be derived as
\begin{subequations}
\begin{equation}\label{eq:muB2}
\textstyle\mu_2^*=\left({P_T+\sum\limits_{k=1}^{r_s}\frac{\sigma^2}{[\boldsymbol{\Lambda}]_{k,k}^2}}\right)\left({r_s\sum\limits_{k=1}^{r_s}\frac{1}{\left(\nu_2^*-[\boldsymbol{\Lambda}]_{k,k}^2\right)\ln2}}\right)^{-1},
\end{equation}
\begin{equation}\label{eq:PAB2}
p_k^{(\rm id)}=\begin{cases}
\frac{\mu_2^*}{\left(\nu_2^*-[\boldsymbol{\Lambda}]_{k,k}^2\right)\ln2} -\frac{\sigma^2}{[\boldsymbol{\Lambda}]_{k,k}^2}, & \text{$k=1,2,\ldots,r_s$}\\
0, & \text{$r_s+1\le k\le r$}
\end{cases}.
\end{equation}
\end{subequations} 
 
The convergence of Algorithm~\ref{Algo:1} to its globally optimal solution is guaranteed due to its generalized convexity property~\cite{Baz,avriel2010generalized}, as proved in Theorem~\ref{th:Gen_Cvx}. However, its speed of convergence depends on the efficiency of the deployed numerical methods for solving the required system of the two non-linear equations \eqref{eq:s1} and \eqref{eq:s2}. Commercial mathematical packages like Matlab or Mathematica provide very efficient solvers for such non-linear systems in the case of existence of a unique solution, as in our considered cases. But the convergence speed of those solvers or conventional subgradient methods~\cite{sboyd_subgradient} depends on the starting point and step sizes.

To characterize the exact number of computations required in achieving a desired level of accuracy with the derived globally optimal solutions, regardless of the starting point and step-sizes fed to the numerical solvers, we next present a simple, yet efficient, 2-D linear search algorithm based on the Golden Section Search (GSS) method~\cite{GS} that provides an effective way of practically implementing Algorithm~\ref{Algo:1}. {We would like to mention that the main steps involved in the global optimization algorithm implemented using Algorithm 2 remain the same as in Algorithm~\ref{Algo:1}. Except that it presents an efficient way of implementing step~\ref{step:I2} of Algorithm~\ref{Algo:1}.}

\subsection{Two-Dimensional (2-D) Linear Search}\label{sec:2D}
As discussed in Section~\ref{sec:problem}, for a known $\rho$, $\mathcal{OP}1$ is a convex optimization problem having a linear objective and convex constraints. Using this property and the small feasible range of $\rho$ given by $0\le\rho_{_\mathrm{LB}}\le\rho\le\rho_{_\mathrm{UB}}\le1$ as derived in Section~\ref{sec:UPS_Ratio_bounds}, we propose to iteratively solve $\mathcal{OP}1$ for a given $\rho$ value till the globally optimal $\left(\mathbf{S^*},\rho^*\right)$ pair is obtained providing the unique maximum received power $P_{R,E}^*$. To traverse over the short value space of $\rho$ we use the GSS method~\cite{GS} that provides fast convergence to the unique root of an equation or a globally optimal solution of a unimodal function. For each feasible $\rho$ value, we substitute into \eqref{eq:s1} and then solve it for the optimal $\nu^*$. As shown in Section~\ref{sec:Multipliers_bounds}, $\nu_{_\mathrm{UB}}-\nu_{_\mathrm{LB}}= 1-\rho\le1$ implying that the search space for the optimal $\nu^*$ is very small. Thus, \eqref{eq:s1} can be solved very efficiently for $\nu^*$ for a given $\rho^*$ value by using the standard one-dimensional (1-D) GSS method or conventional root finding numerical techniques available in most of the commercial mathematical packages.

\subsubsection{Implementation Details}
The detailed algorithmic steps for the proposed 2-D GSS solution are summarized in Algorithm~\ref{Algo:2}. This algorithm includes two linear searches. An outer search aiming at finding $\rho^*$ and an inner one to seek for $\nu^*$ for each given $\rho$ value. Due to the implicit consideration of the constraint $({\rm C3})$, obtaining $\nu^*$ for a given $\rho^*$ involves solving \eqref{eq:s1} using 1-D GSS for at most $r_w\le r$ times, while considering positive power allocation over the $k$ best gain eigenchannels with $k=1,2,\ldots,r$.

\begin{algorithm}[!htp]
	
\small{\caption{\small Global Optimization Algorithm based on the 2-D Golden Section Search}\label{Algo:2}
	\begin{algorithmic}[1]
		\Require $\mathbf{H}$, $\sigma^2$, $P_T$, ${R}$, and acceptable tolerance $\xi\ll1$
		\Ensure Maximized received power $P_{R,E}^*$ for EH along with optimal $\mathbf{S^*}$ and $\rho^*$
		\State Follow steps~\ref{step:in1},~\ref{step:in2}, and~\ref{step:in3} of Algorithm~\ref{Algo:1} for initialization
		\If{${R}>{R_{\max}}$} step~\ref{step:in6} of Algorithm~\ref{Algo:1} \Comment{Infeasible case}
		\ElsIf{${R}\le{R_{\mathrm{th}}}\le{R_{\max}}$} step~\ref{step:I1} of Algorithm~\ref{Algo:1} \Comment{Energy Beamforming mode}
		\Else\Comment{Modified Statistical Multiplexing mode}
		\State Obtain $\rho_{_\mathrm{LB}}$ and $\rho_{_\mathrm{UB}}$ by respectively using \eqref{eq:reb} and \eqref{eq:rho-UB}. Then, set $\rho_p= \rho_{_\mathrm{UB}}-0.618\left(\rho_{_\mathrm{UB}}-\rho_{_\mathrm{LB}}\right)$ \label{stepB:rho1} 
		\State  $r_s=r_w+1$, $\nu_{_\mathrm{LB}}=\rho_p[\boldsymbol{\Lambda}]_{1,1}^2$, and $\nu_{_\mathrm{UB}}=\rho_p[\boldsymbol{\Lambda}]_{1,1}^2+ 1-\rho_p$ \label{stepB:I1}
		\Repeat \mbox{ (Inner loop: Recursion over feasible $\nu$ range)}
		\State Set $r_s=r_s-1$,\, and check if $r_s=1$ to implement step~\ref{step:I1b} of Algorithm~\ref{Algo:1}		
		\State Substitute $\rho^*=\rho_p$ in \eqref{eq:s1}, and solve to obtain $\nu^*\in\left(\nu_{_\mathrm{LB}},\nu_{_\mathrm{UB}}\right)$ using 1-D GSS\label{stepB:I3}
		\State Obtain $\mu^*$ and $p_k^*$ by using steps \ref{step:I3} and~\ref{step:I4} of Algorithm~\ref{Algo:1}\label{stepB:I4}
		\Until{$\left(p_k^*<0\,\,\forall k=1,2,\ldots,r_s\right)$}
		\State Set $P_{R,E}^*=\rho^*\sum_{k=1}^{r_s}p_k^*\,[\boldsymbol{\Lambda}]_{k,k}^2$\label{stepB:I2}
		\State Set $P_{R,p}=P_{R,E}^*$, $\mu_p=\mu^*$, $\nu_p=\nu^*$,  and $\rho_q= \rho_{_\mathrm{LB}}+0.618\left(\rho_{_\mathrm{UB}}-\rho_{_\mathrm{LB}}\right)$ 
		\State \parbox[t]{\dimexpr\linewidth-\algorithmicindent\relax}{Repeat steps~\ref{stepB:I1} to~\ref{stepB:I2} with $\rho_p$ being replaced by $\rho_q$ in these steps to obtain $P_{R,E}^*,\,p_k^*,\,\mu^*,$ and $\nu^*$ for $\rho^*=\rho_q$\strut}
		\State Set $P_{R,q}=P_{R,E}^*,$\;$\mu_q=\mu^*,$\;$\nu_q=\nu^*$\;$ c=0,$ and $\Delta_{\rho}=\rho_{_\mathrm{UB}}-\rho_{_\mathrm{LB}}$\Comment{Here $c$ stores iteration count}
		\While{$\Delta_{\rho}>\xi$} (Outer loop: Recursion over feasible $\rho$)\Comment{Implements conventional 1-D GSS over $\rho$}
		\If{$P_{R,p}\ge P_{R,q}$} 
		\State Set $\rho_{_\mathrm{UB}}=\rho_q$,\,  $\rho_q=\rho_p$,\, $P_{R,q}=P_{R,p}$,\, and $\rho_p= \rho_{_\mathrm{UB}}-0.618\left(\rho_{_\mathrm{UB}}-\rho_{_\mathrm{LB}}\right)$
		\State Repeat steps~\ref{stepB:I1} to~\ref{stepB:I2} with $\rho^*=\rho_p$  and set the results as $P_{R,p}=P_{R,E}^*$, $\mu_p=\mu^*$, $\nu_p=\nu^*$
		\Else 
		\State Set $\rho_{_\mathrm{LB}}=\rho_p$,\, $\rho_p=\rho_q$,\, $P_{R,p}=P_{R,q}$,\, and $\rho_q= \rho_{_\mathrm{LB}}+0.618\left(\rho_{_\mathrm{UB}}-\rho_{_\mathrm{LB}}\right)$ 
		\State Repeat steps~\ref{stepB:I1} to~\ref{stepB:I2} with $\rho_p$ replaced by $\rho_q$ and to obtain $P_{R,q}=P_{R,E}^*$, $\mu_q=\mu^*$,  $\nu_q=\nu^*$
		\EndIf
		\State Set  $c=c+1$ and $\Delta_{\rho}=\rho_{_\mathrm{UB}}-\rho_{_\mathrm{LB}}$
		\EndWhile 
		\EndIf 
	\end{algorithmic}} 
\end{algorithm}  
	
Algorithm~\ref{Algo:2} can also be slightly modified to be used for obtaining the globally optimal solution for $\mathcal{OP}2$. In this case, due to ideal reception, the outer GSS over the feasible $\rho$ values has to be removed and we only need to perform a 1-D GSS for $\nu_2^*$ over its feasible value range $[\boldsymbol{\Lambda}]_{1,1}^2\le\nu_2^*\le[\boldsymbol{\Lambda}]_{1,1}^2+1$. Therefore, for $\mathcal{OP}2$ we need to consider steps 1--\ref{stepB:I2} of Algorithm~\ref{Algo:2}, excluding the initialization  step~\ref{stepB:rho1}, and updating steps~\ref{stepB:I1},~\ref{stepB:I3},~\ref{stepB:I4}, and~\ref{stepB:I2}. Particularly, the bounds are given by $\nu_{2_\mathrm{LB}}=[\boldsymbol{\Lambda}]_{1,1}^2$ and $\nu_{2_\mathrm{UB}}=[\boldsymbol{\Lambda}]_{1,1}^2+1$ in step~\ref{stepB:I1}. In step~\ref{stepB:I3}, we need to solve \eqref{eq:s1-I} to find optimal $\nu_2^*$ for ${R}>{R_{\mathrm{th}}}$. This $\nu_2^*$ value will then be used in step~\ref{stepB:I4} to obtain the optimal $\mu_2^*$ and $p_k^{(\rm id)}$'s by substituting $\nu_2^*$ in \eqref{eq:muB2} and \eqref{eq:PAB2}. Lastly, we need to set $\rho^*=1$ in step~\ref{stepB:I2}.

\subsubsection{Complexity Analysis}
Suppose that we want to calculate $\rho^*$ and $\nu^*$ of $\mathcal{OP}1$ or $\nu_2^*$ of $\mathcal{OP}2$ through Algorithm~\ref{Algo:2} so as to be close up to an acceptable tolerance $\xi\ll1$ to their globally optimal solutions. As seen from Algorithm~\ref{Algo:2}, the search space interval after each GSS iteration reduces by a factor of $0.618$~\cite[Chap$.$ 2.5]{GS}. This value combined with the unity maximum search length for $\rho^*$ and $\nu^*$ gives the number of iterations $c^* = \ceil[\Big]{\frac{\ln\left(\xi\right)}{\ln\left(0.618\right)}}+1$ that are required to ensure that the numerical error is less than $\xi$. For example, $\xi=10^{-3}$ results in $c^*=16$. Note that $c^*$ is a logarithmic function of $\xi$ and is independent of $N_T$, $N_R$, and $r$. As each computation

\noindent in GSS iteration for finding $\rho^*$ involves an inner GSS for computing $\nu^*$, which is repeated for at most $r_w$ runs, the total number of iterations required for finding the globally optimal solution of $\mathcal{OP}1$ within an acceptable tolerance $\xi$ is given by $c_1^*\le r_w\,c^*\left(c^*+1\right)$. Since the number of function computations in GSS is one more than the number of iterations and $r_s\le r_w \le r$ from Lemma~\ref{lem:rnk}, the total number of computations involved in solving $\mathcal{OP}1$ are bounded by the value $r\left(\ceil[\Big]{\frac{\ln\left(\xi\right)}{\ln\left(0.618\right)}}+2\right)^2$. Hence, the computational complexity of Algorithm~\ref{Algo:2} is $\mathcal{O}\left(r\right)$, i$.$e$.$, linear in $r$. This complexity witnesses the significance of Algorithm~\ref{Algo:2} over Algorithm~\ref{Algo:1}. Instead of directly implementing commercial numerical solvers or subgradient or ellipsoid methods~\cite{sboyd_subgradient} for Algorithm~\ref{Algo:1}, we use the 2-D GSS method as outlined in Algorithm~\ref{Algo:2}. 

Regarding the required number of iterations $c_2^*$ for finding the globally optimal solution of $\mathcal{OP}2$ it must hold $c_2^*\le r_w\,c^*\le r\,c^*$. As a result, the computational complexity of the modified Algorithm~\ref{Algo:2} for $\mathcal{OP}2$ is $\mathcal{O}\left(r\left(\ceil[\Big]{\frac{\ln\left(\xi\right)}{\ln\left(0.618\right)}} +1\right)\right)=\mathcal{O}\left(r\right)$, i.e., it's also linear in $r$.

\subsubsection{High SNR Approximation}\label{sec:asymimp}
Recalling Remark~\ref{rem:asym} in Section~\ref{sec:cf-asym} holding for high SNR values and focusing on equations~\eqref{eq:PA-a-I2} and \eqref{eq:PA-a-I3} for ${R}>{R_{\mathrm{th}}}$, it becomes apparent that, since $p_k^{(\rm id,a)}>0$ $\forall k=1,2,\ldots,r$, then even with the implicit consideration of constraint $({\rm C3})$ one does not need to repeatedly solve the 1-D GSS over $\nu_2^*$ for at most $r$ times. Thus, we only need to find $\beta\in(0,1)$ from the following equation using the 1-D GSS method:
\begin{align}\label{eq:PA-a-I5}
\textstyle{\frac{P_T}{\sigma^2}}\left(\prod\limits_{k=1}^{r}\frac{\beta\,[\boldsymbol{\Lambda}]_{k,k}^2}{\beta+[\boldsymbol{\Lambda}]_{1,1}^2-[\boldsymbol{\Lambda}]_{k,k}^2}\right)^{\frac{1}{r}}&\,=2^{\frac{R}{r}}\,{\textstyle\sum\limits_{k=1}^{r}\left(\frac{\beta}{\beta+[\boldsymbol{\Lambda}]_{1,1}^2-[\boldsymbol{\Lambda}]_{k,k}^2}\right)}.
\end{align} 
The computational complexity of finding the globally optimal solution of $\mathcal{OP}2$ for high SNR values is therefore $\mathcal{O}\left(\ceil[\Big]{\frac{\ln\left(\xi\right)}{\ln\left(0.618\right)}}+1\right)=\mathcal{O}\left(1\right)$, i.e., constant or independent of $r$.

\section{Numerical Results and Discussion}\label{sec:results} 
In this section, we numerically evaluate the performance of the proposed joint TX precoding and RX UPS splitting design, and investigate the impact of various system parameters on its achievable rate-energy trade off. Unless otherwise stated, we set $\sigma^2=\left\{-100,-70\right\}$dBm by considering noise spectral density of $-175$ dBm/Hz as well as $P_T=10$W, and $\xi=10^{-4}$. Furthermore, we model $\mathbf{H}$ as $\mathbf{H}=\big\{\theta h_{ij}\,\big| 1\le i,j\le N\big\}$ with $N\triangleq N_R=N_T=\left\{2,4\right\}$, where $\theta=\left\{0.1,0.05\right\}$ models the distance dependent propagation losses and $h_{ij}$'s are ZMCSCG random variables with unit variance. {With this definition, the average channel power gain is given by  $\theta^2=a \,d^{-n}$, where $a$ is the propagation loss constant, $n$ is the path loss exponent, and $d$ is the TX-to-RX distance. So, for $a=0.1$ and $n=2$, $\theta=0.1$ represents that $d=3.16$ m. Whereas this separation becomes twice, i.e., $d=6.32$ m, for $\theta=0.05$.} We assume unit transmission block duration, thus, we use the terms `received energy' and `received power' interchangeably. All performance results have been generated after averaging over $10^3$ independent channel realizations. For obtaining the globally optimal $\mathbf{S}^*$ and $\rho^*$ with the proposed design we have simulated Algorithm~\ref{Algo:2}. 
\begin{figure}[!t]
\centering
\subfigure[$2\times 2$ MIMO system.]{{\includegraphics[width=2.8in]{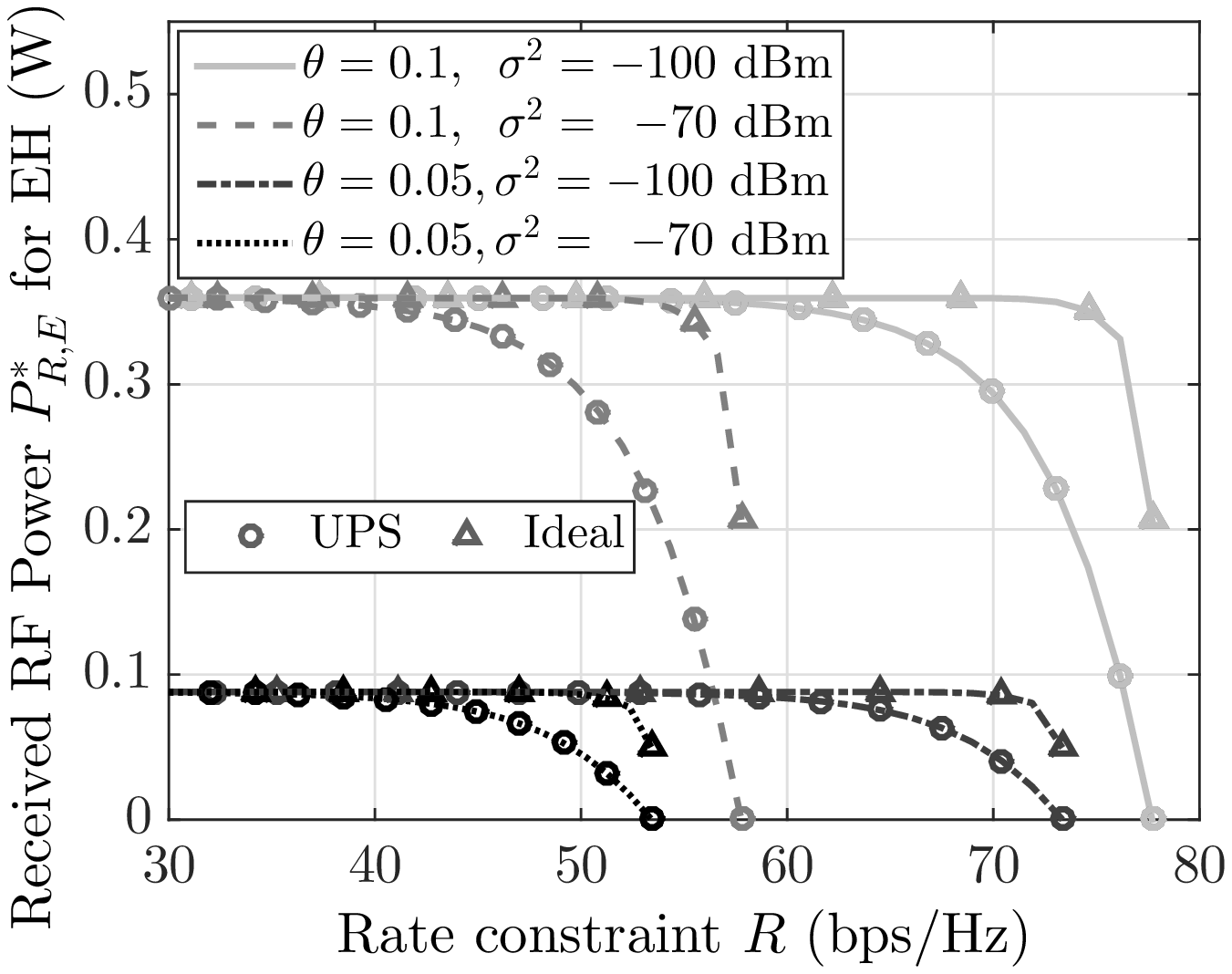} }}\qquad
\subfigure[$4\times 4$ MIMO system.]
{{\includegraphics[width=2.8in]{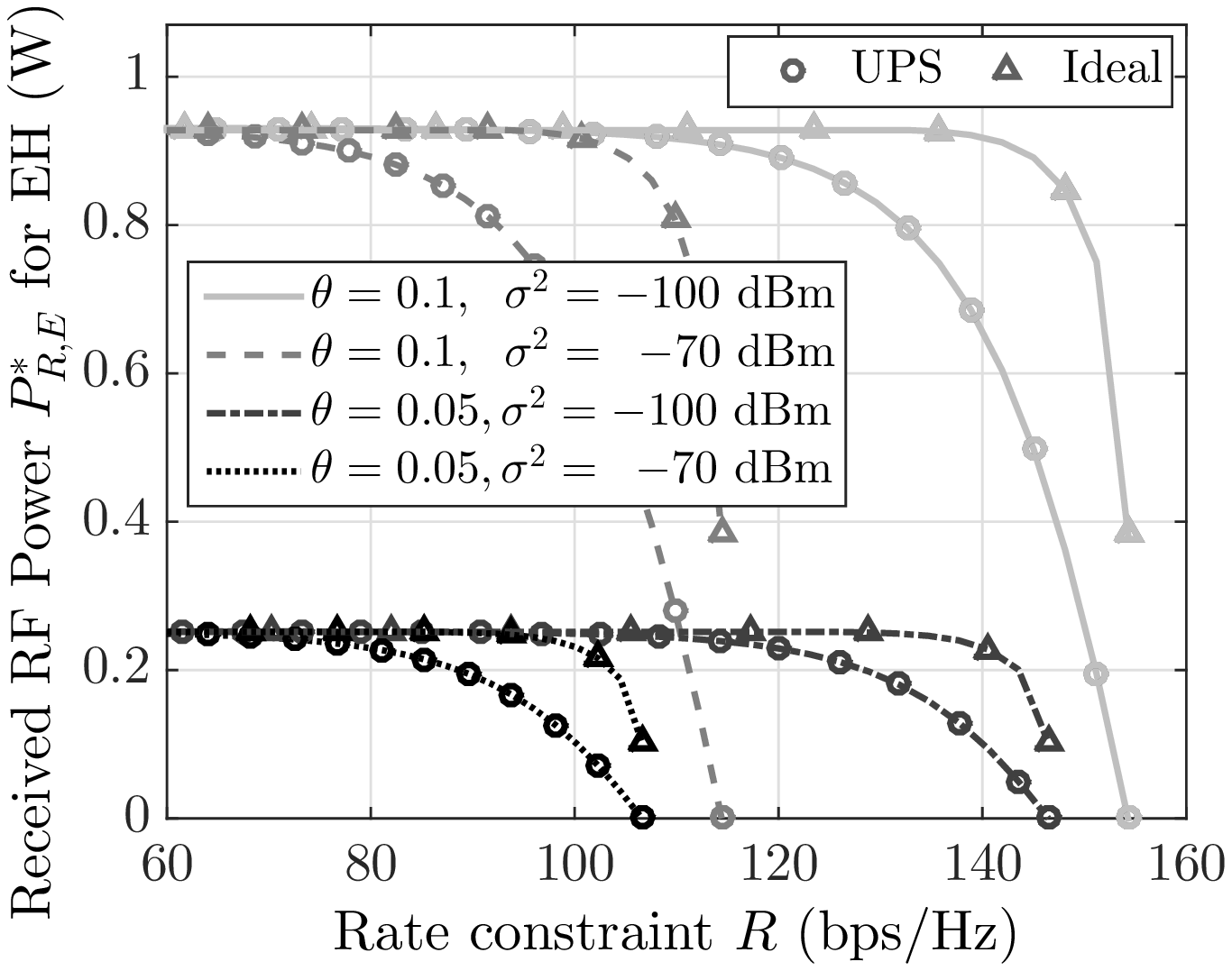} }}
\caption{\small Variation of the rate-energy trade off for $P_T=10$W and different values for $N$, $\theta$, and $\sigma^2$.}
    \label{fig:tradeoff} 
\end{figure}

\begin{figure}[!t]
	\centering
	\subfigure[{RF-to-DC rectification efficiency variation.}]{{\includegraphics[width=2.7in]{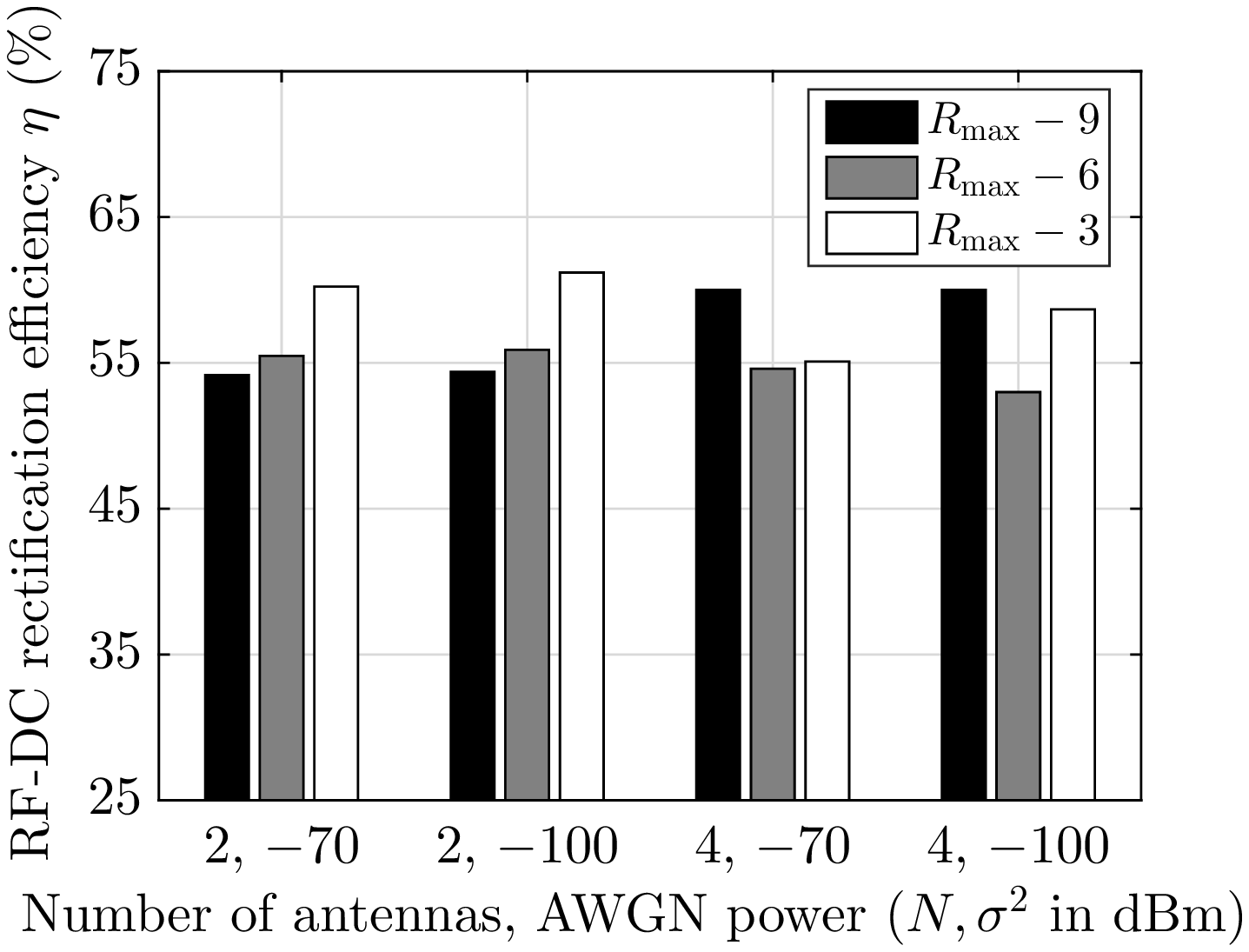} }}\, 
	\subfigure[{Harvested DC power variation.}]
	{{\includegraphics[width=2.7in]{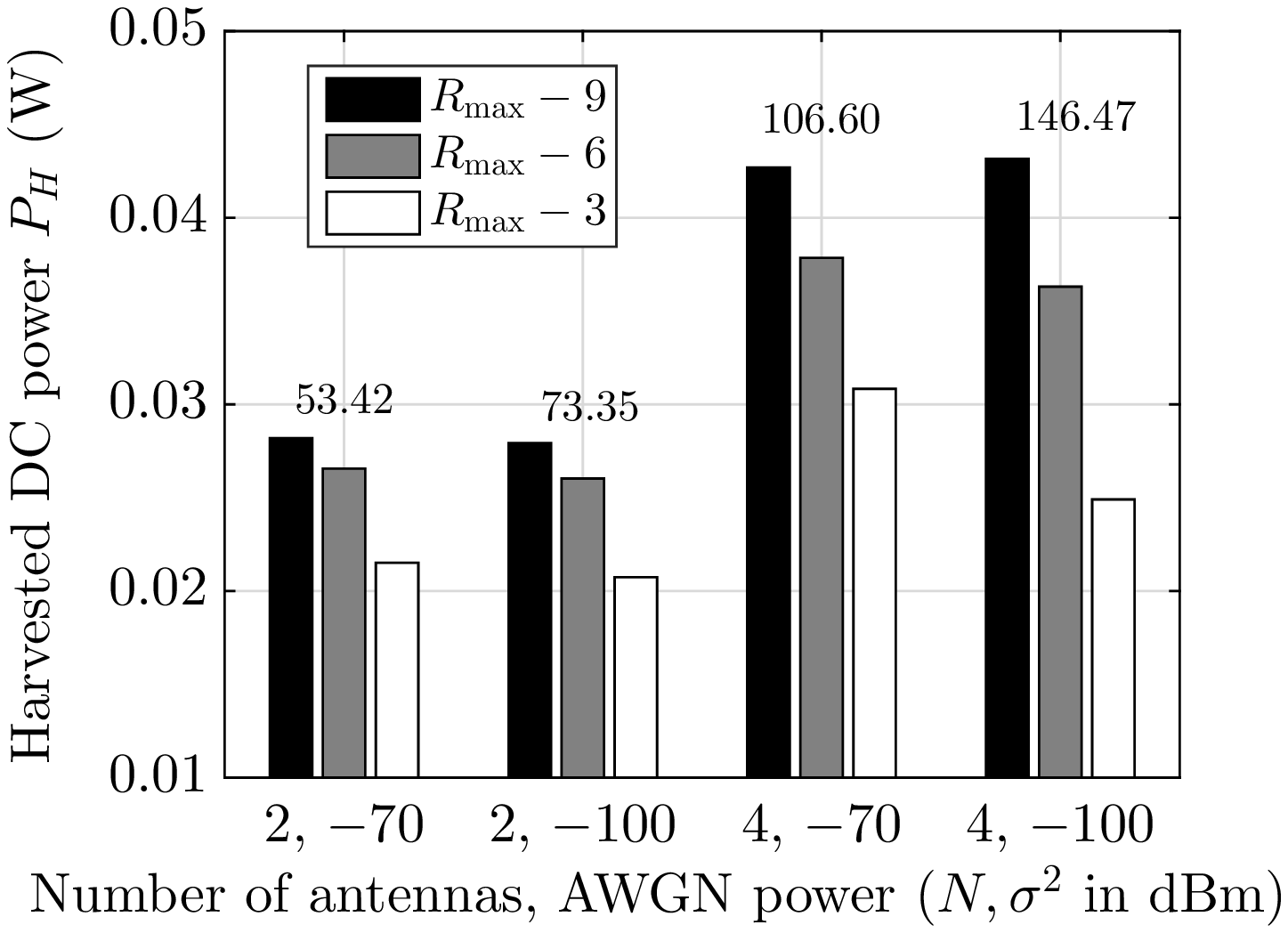} }} 
	\caption{\small {Variation of rectification efficiency and harvested  power for Powercast RF EH circuit~\cite{Powercast}  with rate $R$ near $R_{\max}$ for $P_T=10$W $\theta=0.05$, and different values for $N$ and $\sigma^2$. Also, $R_{\max}$ is mentioned over the bars in (b).}} 
	\label{fig:Eta} 
\end{figure}
We consider $2\times 2$ and $4\times 4$ MIMO systems in Fig$.$~\ref{fig:tradeoff} with both ideal and UPS reception and illustrate the rate-energy trade off for our proposed designs for different values for the propagation losses and noise variance parameters. As expected, our solution for $\mathcal{OP}2$ with ideal reception outperforms that of $\mathcal{OP}1$ that considers practical UPS reception. It is also obvious that increasing $N$ improves the rate-energy trade off. This happens because both beamforming and multiplexing gains improve as $N$ gets larger. Lesser noisy systems, when $\sigma^2$ decreases, and better channel conditions with increasing $\theta$ result in better trade off and enable higher achievable rates. The maximum achievable rate $R_{\max}$ in bps/Hz for the considered four cases $\left(\theta,\sigma^2\right)=\left\{\left(0.05,-70\text{dBm}\right),\left(0.1,-70\text{dBm}\right),\left(0.05,-100\text{dBm}\right),\left(0.1,-100\text{dBm}\right)\right\}$ is given by $\left\{53.42,57.77,73.35,77.7\right\}$ for $N=2$ and $\left\{106.60,114.42,146.47,154.28\right\}$ for $N=4$. In addition, the average value of $R_{\mathrm{th}}$ in bps/Hz for these cases is given by $\left\{17.27,18.72,26.02,27.98\right\}$ for $N=2$ and $\left\{17.46,19.05,26.92,28.73\right\}$ for $N=4$. When the rate requirement $R$ is below $R_{\mathrm{th}}$, the maximum received RF power $P_{R,E}^*$ for EH is achieved with TX energy beamforming. However, as $R$ increases and becomes substantially larger than $R_{\mathrm{th}}$, $P_{R,E}^*$ decreases till reaching a minimum value. For the latter cases, TX spatial multiplexing is adopted to achieve $R$ and any remaining received power is used for EH. {Further, with $\theta$ decreasing as $\left\lbrace 0.1,0.01,0.001\right\rbrace$, the corresponding $R_{\max}$ varies as   
$\{ 77.70,64.06,$ $50.76\}$ bps/Hz for $N=2$ and $\left\lbrace 154.28,127.89,101.33\right\rbrace$ bps/Hz for $N=4$. Whereas, the corresponding maximum achievable RF power $P_{R,E}^*$ for EH with $R=0$ bps/Hz varies as $\{0.36\,\text{W}, 3.52\,\text{mW}, 35.74\,\mu\text{W}\}$ for $N=2$ and $\{ 0.93\,\text{W}, 10.09\,\text{mW},$ $97.22 \,\mu\text{W}\}$ for $N=4$ MIMO SWIPT systems.}

{Now considering Powercast RF EH circuit~\cite{Powercast}, we investigate the impact of the non-linear rectification efficiency $\eta$ on the optimized harvested DC power $P_H$ with varying rate requirements close to  $R_{\max}$ because in this regime the corresponding $P_{R,E}^*$ decreases sharply as shown in Fig.~\ref{fig:tradeoff}. For each of the four cases  of varying $N$ and $\sigma^2$ as plotted in Fig.~\ref{fig:Eta}, though $\eta$ does not follow any trend (increasing for first two cases and decreasing then increasing for the next two), the optimized harvested DC power $P_H$ is monotonically decreasing with increasing $R$ from $R=R_{\max}-9$ bps/Hz to $R=R_{\max}-3$ bps/Hz, because this increase in rate $R$ results in a lower $P_{R,E}^*$. So, this monotonic trend of optimized $P_H$ in $P_{R,E}^*$ as depicted via Fig.~\ref{fig:Eta} numerically corroborates the discussion with respect to the claim made in Proposition 1 and  the RF EH characteristics as plotted in Fig.~\ref{fig:NonLinear-RFEH}.}


\begin{figure*}[!t]	
	\begin{minipage}{.48\textwidth}	
		\centering
		\centering
		{{\includegraphics[width=2.8in]{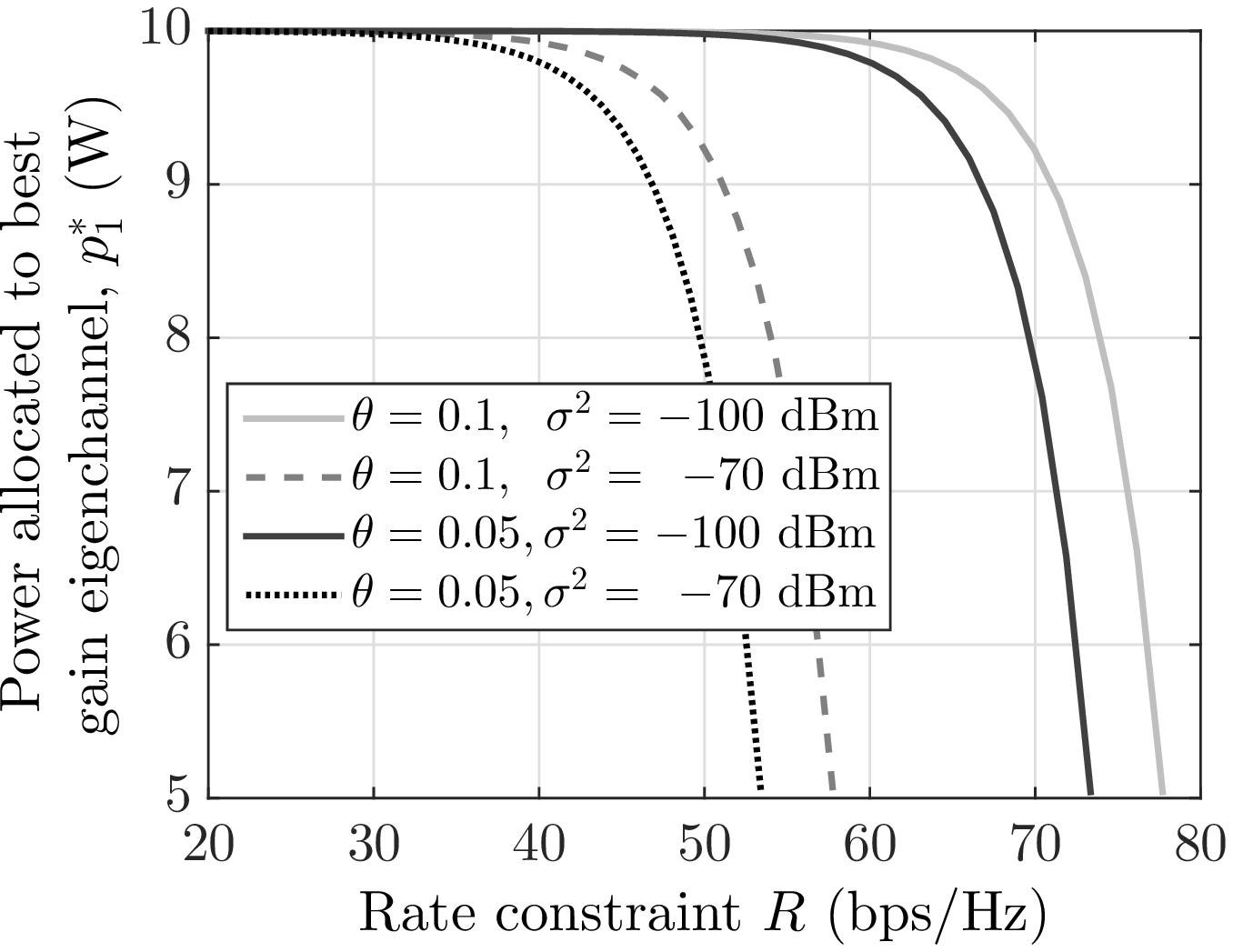} }}
		\caption{\small Variation of the optimal power allocation $p_1^*$ of the best gain eigenchannel of a $2\times 2$ MIMO system as a function of the rate constraint $R$ for $P_T=10$W.}    \label{fig:PA2x2}  
	\end{minipage}\quad
	\begin{minipage}{.48\textwidth}		
		\centering
		{{\includegraphics[width=2.8in]{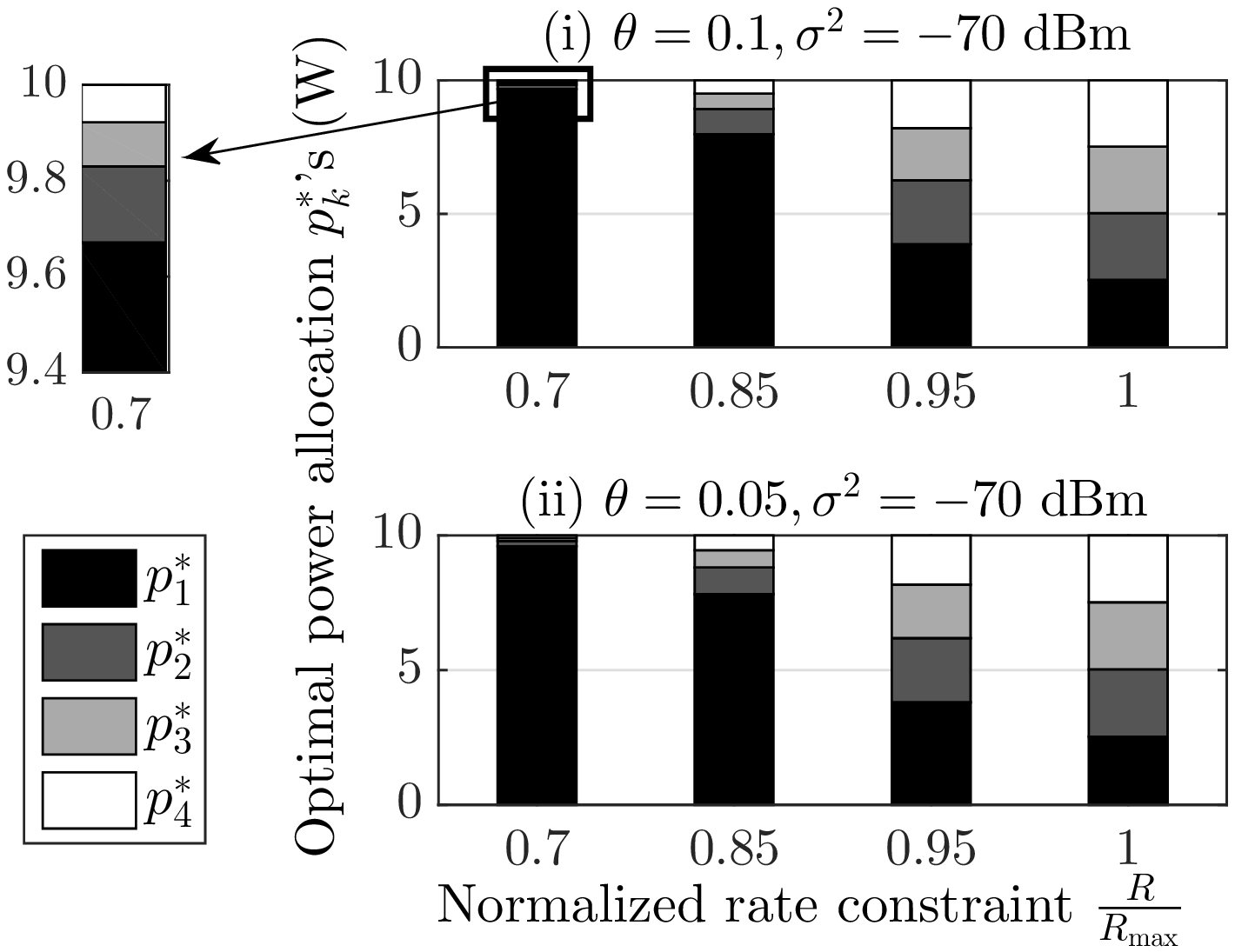} }}
		\caption{\small {Variation of optimal power allocation $p_1^*$, $p_2^*$, $p_3^*$, $p_4^*$ in a $4\times 4$ MIMO system as a function of normalized rate constraint $\frac{R}{R_{\max}}$ for $P_T=10$W and $\sigma^2=-70$dBm.}}
		\label{fig:PA4x4}
	\end{minipage}	
\end{figure*}

The variation of the optimal power allocation with the proposed joint design for $\mathcal{OP}1$ is depicted in Figs$.$~\ref{fig:PA2x2} and~\ref{fig:PA4x4} for $2\times 2$ and $4\times 4$ MIMO systems, respectively, as a function of the rate constraint $R$. Particularly, Fig$.$~\ref{fig:PA2x2} illustrates the optimal power allocation $p_1^*$ over the best gain eigenchannel, while the optimal power allocation $p_1^*$, $p_2^*$, $p_3^*$, and $p_4^*$ over the $r=4$ available eigenchannels is demonstrated in Fig$.$~\ref{fig:PA4x4}. As shown, $p_1^*$ monotonically decreases from $p_1^*{\,\approxeq\,} P_T$ (this happens for $R\le R_{\mathrm{th}}$ where TX energy beamforming is adopted) to the equal power allocation $p_1^*{\,\approxeq\,} p_2^*{\,\approxeq\,}\frac{P_T}{2}$ (for large $R=R_{\max}$, TX spatial multiplexing is used).  {As from \eqref{eq:PA}, $p_1^*\ge p_2^*$, we note that with $P_T=10$W for $N=2$, $p_1^*\ge 5$ W in Fig.~\ref{fig:PA2x2}.} Similar trend to the power allocation of Fig$.$~\ref{fig:PA2x2} is observed in Fig.~\ref{fig:PA4x4}.  It can be observed that, for the plotted normalized rate constraint range, most of $P_T$ is allocated to the best gain eigenchannel in order to perform TX energy beamforming, while the remaining power is allocated to the rest eigenchannels in order to meet the rate requirement $R$ with spatial multiplexing.  

\begin{figure*}[!t]	
	\begin{minipage}{.48\textwidth}	
	\centering
	{{\includegraphics[width=2.8in]{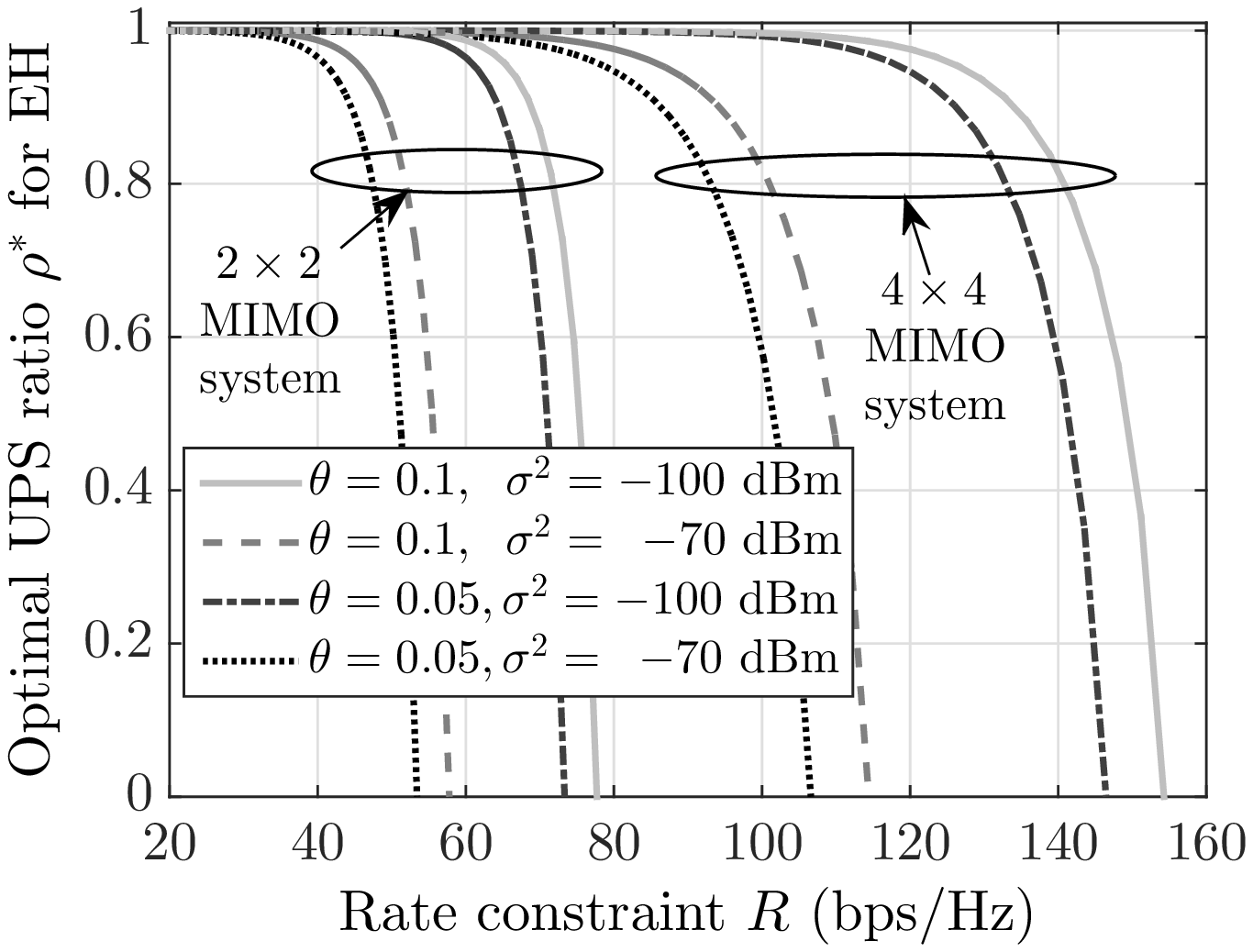} }}
	\caption{\small Variation of the optimal UPS ratio $\rho^*$ for $\mathcal{OP}1$ versus the rate constraint $R$ for $P_T=10$W and different values for $N$, $\theta$, and $\sigma^2$.} \label{fig:UPS}
	\end{minipage}\quad
	\begin{minipage}{.48\textwidth}		
		\centering
		{{\includegraphics[width=2.48in]{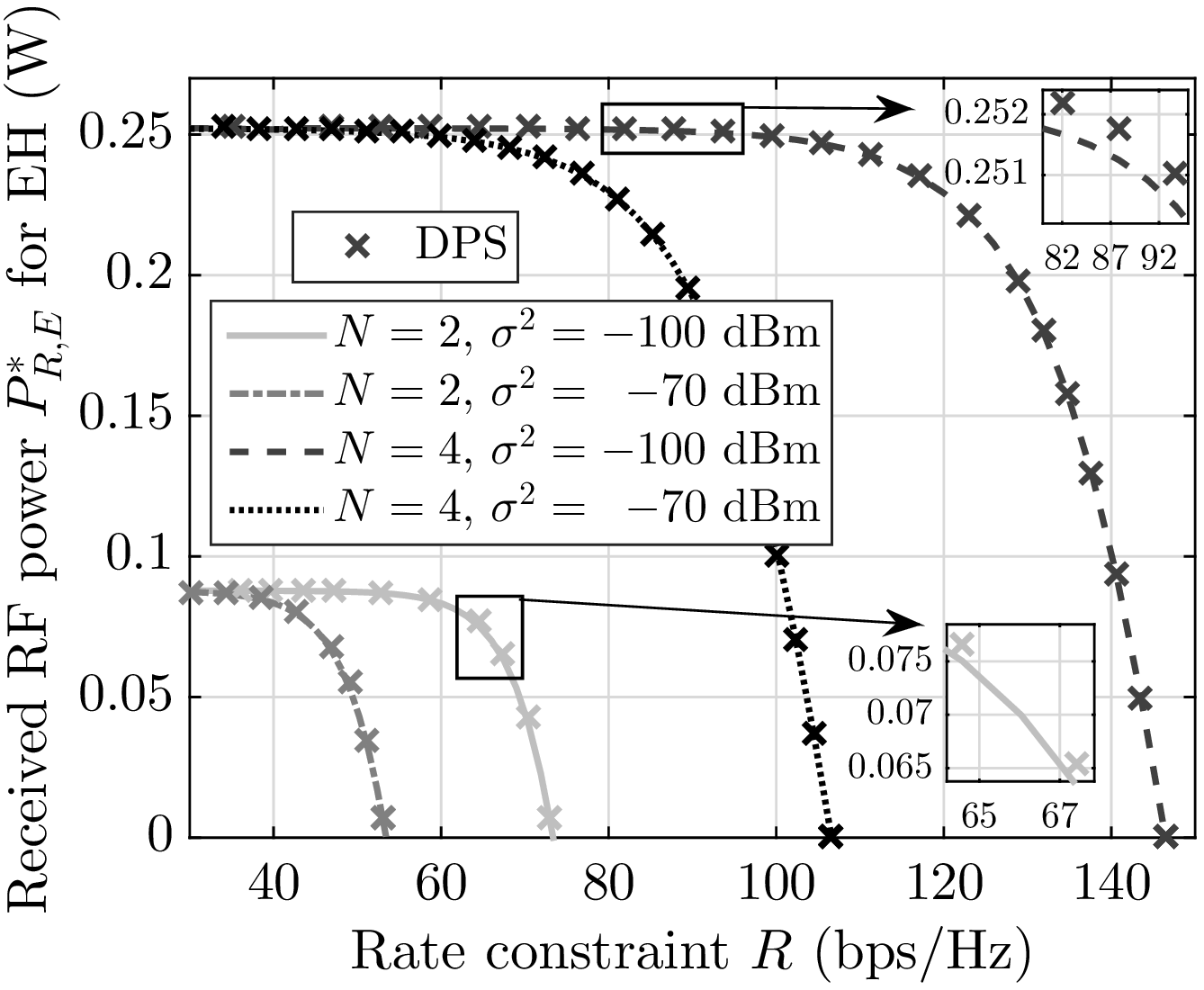} }} 
		\caption{\small Comparison against the numerically obtained optimized DPS (plotted using '$\times$' markers) for $P_T=10$W, $\theta=0.05$, and different values for $N$ and $\sigma^2$.} 
		\label{fig:DPS}
	\end{minipage}	
\end{figure*}

In Fig.~\ref{fig:UPS}, the optimal UPS ratio $\rho^*$ is plotted versus $R$ for $2\times 2$ and $4\times 4$ MIMO systems. It is shown that $\rho^*$ monotonically decreases with increasing $R$ in order to ensure that sufficient fraction of the received RF power 
is used for ID, thus, to satisfy the rate requirement. Lower $\sigma^2$, larger $N$ or equivalently $r$, and higher $\theta$ result in meeting $R$ with lower fraction $1-\rho$ of the received RF power dedicated for ID. Thus, for these cases for a given $R$, larger portion of the received RF power can be used for EH.

\textcolor{black}{We henceforth compare the considered UPS RX operation against the more generic Dynamic PS (DPS) design, according to which each antenna has a different PS value. Since replacing DPS in our formulation results in a non-convex problem, we obtain the optimal PS ratios for the $N$ RX antennas from a $N$-dimensional linear search over the $N$ PS ratios $\rho_1,\rho_2,\ldots,\rho_N$ to select the best possible $N$-tuple. In Fig.~\ref{fig:DPS}, we plot $P_{R,E}^*$ for both UPS and DPS RX designs for $2\times 2$ and $4\times 4$ MIMO systems with $\theta=0.05$ and varying $\sigma^2$. As seen from all cases, the performance of optimized DPS is closely followed by the optimized UPS with an average performance degradation of less than $0.9$mW for $N=2$ and $2.1$mW for $N=4$. This happens because the average deviation of all PS ratios in the DPS design from the UPS ratio $\rho$ is less than $0.001$. A similar observation regarding the near-optimal UPS performance was also reported in~\cite{DPS} for $N_T=1$ at TX. This comparison study corroborates that the adoption of UPS instead of DPS that incurs very high implementation complexity without yielding relatively large gains.}

\begin{figure}[!t]
	\centering
	\subfigure[Optimal Lagrange multiplier $\mu^*$.]{{\includegraphics[width=2.8in]{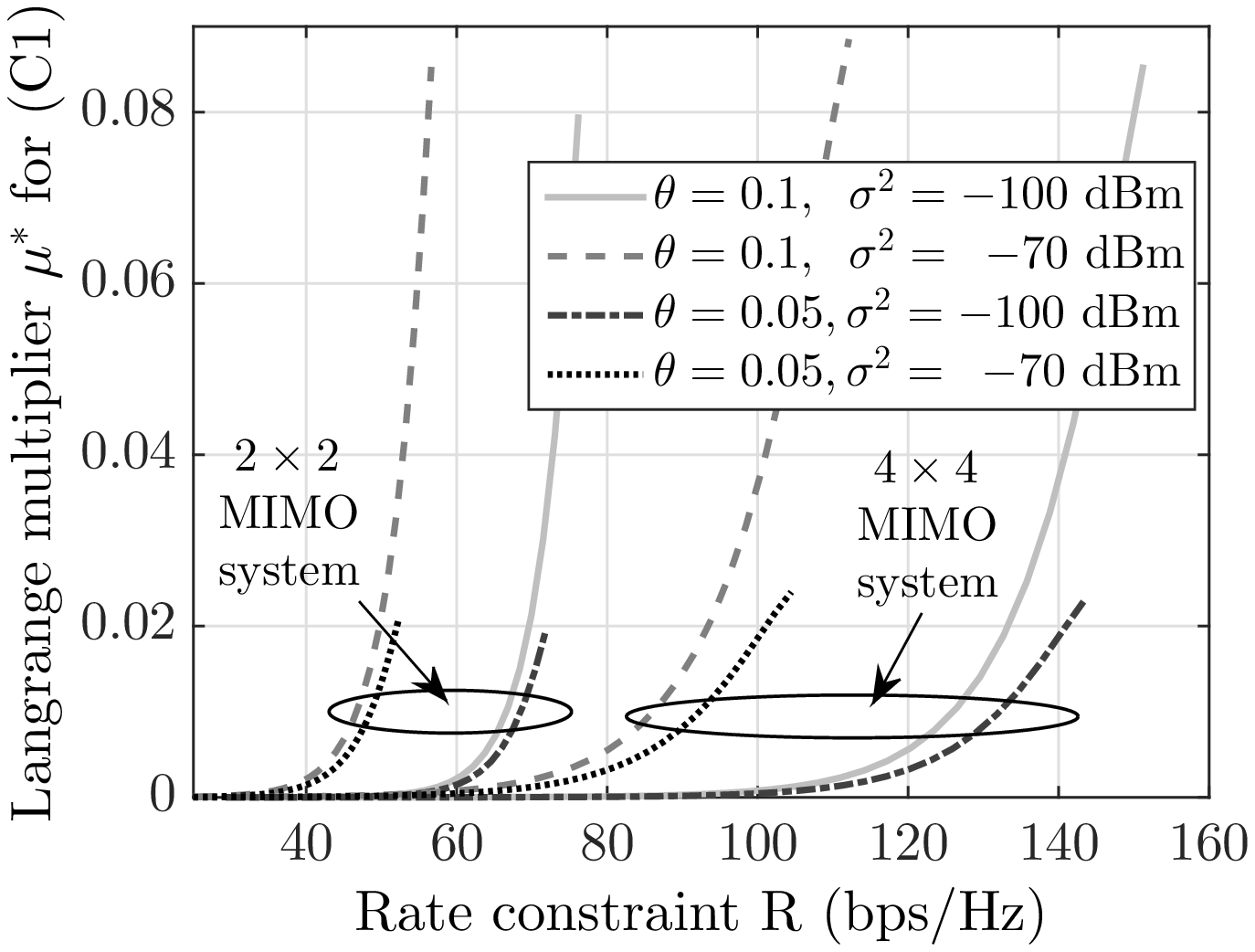} }}\qquad
	\subfigure[Optimal Lagrange multiplier $\nu^*$.]
	{{\includegraphics[width=2.8in]{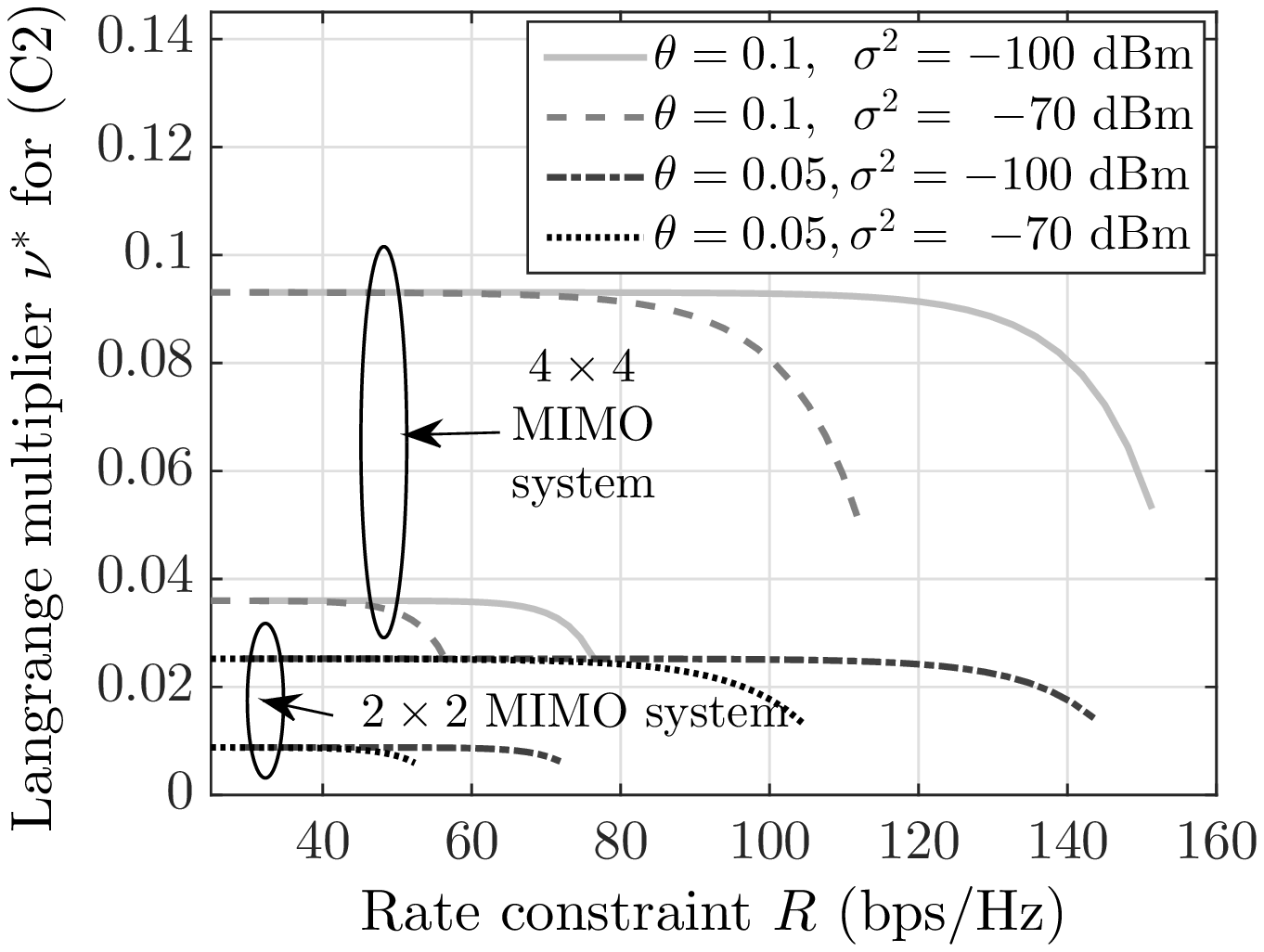} }}
	\caption{\small Variation of optimal Lagrange multipliers for $\mathcal{OP}1$ for $R>R_{\mathrm{th}}$, $P_T=10$W, and varying $N$, $\theta$, and $\sigma^2$.}
	\label{fig:LM} 
\end{figure}
The Lagrange multipliers $\mu^*$ and $\nu^*$ in $\mathcal{OP}1$ are available in closed-form as shown in \eqref{eq:cfmu} and \eqref{eq:cfnu}, respectively, for $R\le R_{\mathrm{th}}$, i$.$e$.$, when energy beamforming is adopted as our TX precoding design. However, one needs to solve a system of non-linear equation for these multipliers, as described in Section~\ref{sec:red1}, for $R>R_{\mathrm{th}}$. In Fig$.$~\ref{fig:LM}, we plot the variation of $\mu^*$ and $\nu^*$ in $\mathcal{OP}1$ for $R>R_{\mathrm{th}}$. As shown, $\mu^*$ and $\nu^*$  monotonically increase and decrease, respectively, with increasing $R$. The average value for $[\boldsymbol{\Lambda}]_{1,1}^2$ for the considered pair values $\left(\theta,N\right)=\left\{\left(0.1,2\right),\left(0.05,2\right),\left(0.1,4\right),\left(0.05,4\right)\right\}$ is $\left\{0.036,0.009,0.093,0.025\right\}$, and it is evident from Fig$.$~\ref{fig:LM}(b) that $\nu^*$ is very close to its lower bound given by $\nu_{_\mathrm{LB}}=\rho^*[\boldsymbol{\Lambda}]_{1,1}^2$. Also, Fig$.$~\ref{fig:LM}(a) showcases that the range of $\mu^*$ is similarly small to $\nu^*$. These findings corroborate the fast convergence of Algorithm~\ref{Algo:2} that exploits the short search space of $\nu^*$ in the solution of $\mathcal{OP}1$ or $\nu_2^*$ in $\mathcal{OP}2$. 
\begin{figure}[!t]
\centering
{{\includegraphics[width=6.3in]{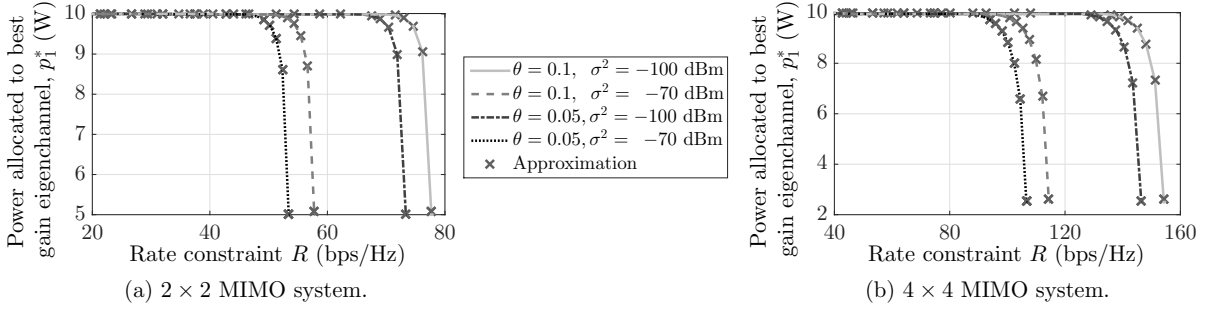} }}
\caption{\small Validating the accuracy of the {proposed high SNR approximation for the globally optimal power allocation $p_1^*$ for $\mathcal{OP}2$} with $P_T=10$W and different values for $N$, $\theta$, and $\sigma^2$.}
    \label{fig:approx}
\end{figure}
Figure~\ref{fig:approx} includes results with the derived tight asymptotic approximation $\mathbf{{S_{\text{id,a}}^*}}$ for the globally optimal solution $\mathbf{S_{\text{id}}^*}$ of $\mathcal{OP}2$ in Section~\ref{sec:cf-asym} using the efficient implementation of Section~\ref{sec:asymimp}. As shown, the results with the TX precoding design $\mathbf{{S_{\text{id,a}}^*}}$ (or $\mathbf{{P_{\text{id,a}}^*}}$), which have been obtained from the solution of the single equation \eqref{eq:PA-a-I5} of $\beta$, match very closely with the results for the globally optimal design $\mathbf{S_{\text{id}}^*}$ (or $\mathbf{P_{\text{id}}^*}$) for $\mathcal{OP}2$ implemented using Algorithm~\ref{Algo:2}.  

\begin{figure}[!t]
	\centering
	\subfigure[$2\times2$  MIMO system.]{{\includegraphics[width=2.8in]{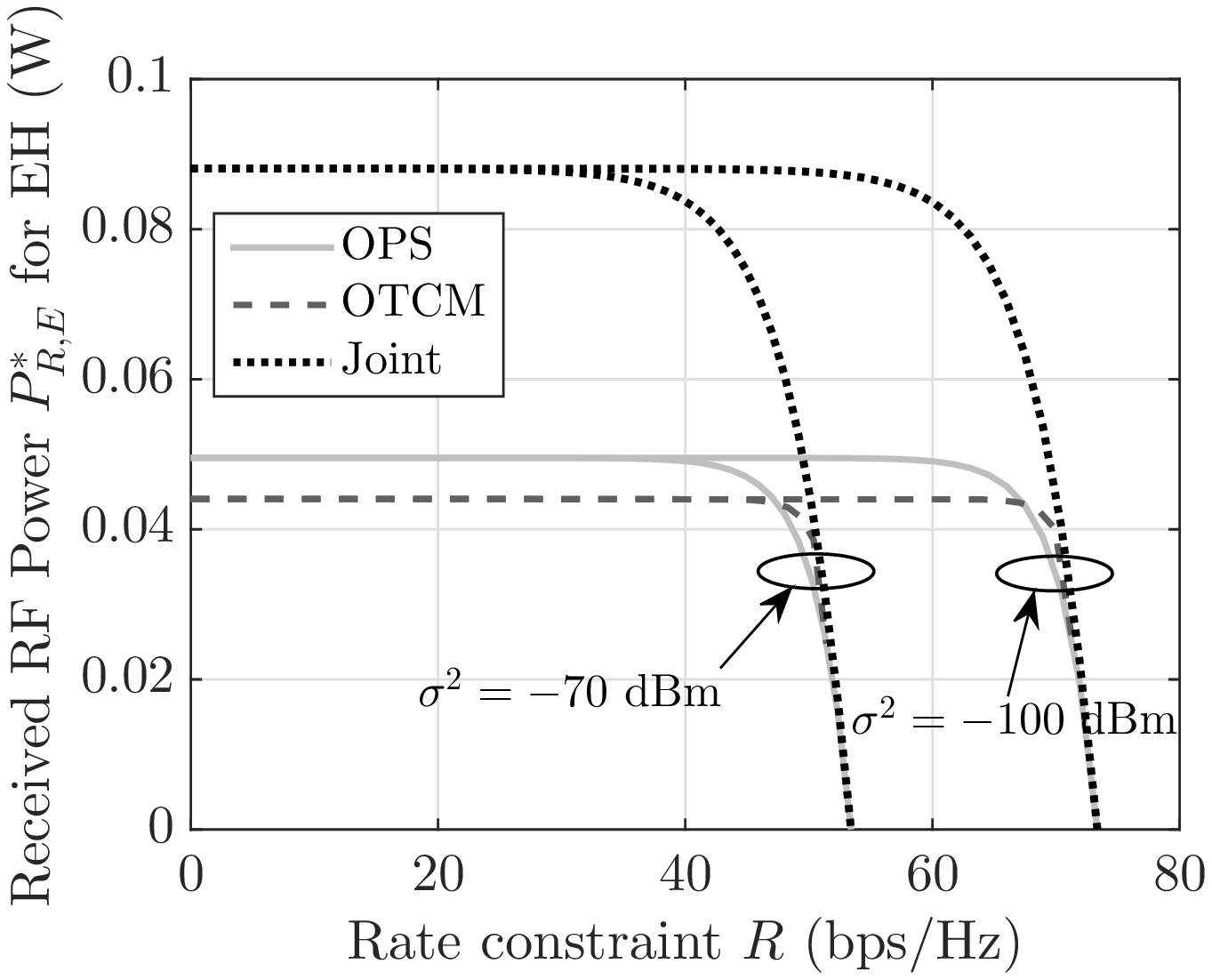} }}
	\qquad\subfigure[$4\times4$  MIMO system.]
	{{\includegraphics[width=2.8in]{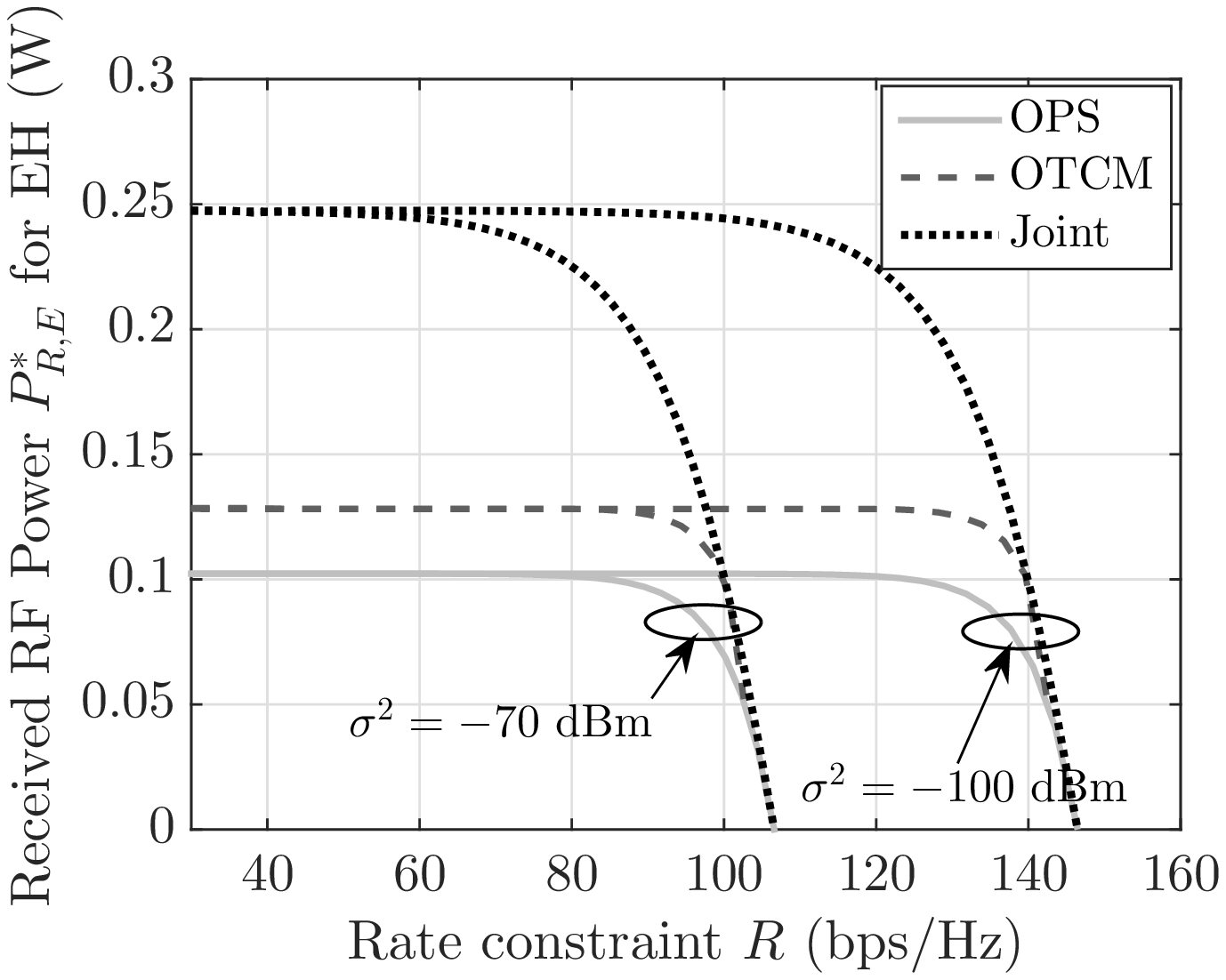} }}
	\caption{\small Comparison of the rate-energy trade off between the proposed joint TX and RX design and the benchmark semi-adaptive schemes OPS and OTCM for $P_T=10$W and $\theta=0.05$ as well as different values for $N$ and $\sigma^2$.}\label{fig:comp}
\end{figure}

\color{black}
We finally present in Fig.~\ref{fig:comp} performance comparison results between the proposed joint TX precoding and RX UPS splitting design, as obtained from the solution of $\mathcal{OP}1$, and the following two benchmark schemes to highlight the importance of our considered joint optimization framework. The first scheme, termed as Optimal TX Covariance Matrix (OTCM), performs optimization of the TX covariance matrix $\mathbf{S}$ for a fixed UPS ratio $\rho=0.5$, and the second scheme, termed as Optimal UPS Ratio (OPS), optimizes $\rho$ for given $\mathbf{S}=\mathbf{S_{_\mathrm{WF}}}$. It is observed that for $2\times2$ MIMO systems, OPS performs better than OTCM, while for $4\times4$ MIMO systems, the converse is true. This happens because OTCM performance improves with increasing $N$ or equivalently $r$. For both $N$ value, the proposed joint TX and RX design provides significant energy gains over OTCM and OPS. Particularly, the performance enhancement for $N=2$ is $71.15$\% and $87.4$\%, respectively, over OPS and OTCM schemes, while this enhancement becomes $127.0$\% and $77.4$\%, respectively, for $N=4$.\color{black}

\section{Conclusions}\label{sec:conclusion} 
In this paper, we investigated EH as an add-on feature in conventional MIMO systems that only requires incorporating UPS functionality at reception side. We particularly considered the problem of jointly designing TX precoding operation and UPS ratio to maximize harvested power, while ensuring that the quality-of-service requirement of the MIMO link is satisfied. By proving the generalized convexity property for a specific reformulation of the harvested power maximization problem, we derived the global jointly optimal TX precoding and RX UPS ratio design. We also presented the globally optimal TX precoding design for ideal reception. Different from recently proposed designs, the solutions of both considered optimization problems with UPS and ideal RXs unveiled that there exists a rate requirement value that determines whether the TX precoding operation is energy beamforming or information spatial multiplexing. We also presented analytical bounds for the key variables of our optimization problem formulation along with tight  practically-motivated high SNR approximations for their optimal solutions. We presented an algorithm for efficiently solving the KKT conditions for the considered problem for which we designed a linear complexity implementation that is based on 2-D GSS. \textcolor{black}{Its complexity was shown to be independent of the number of transceiver antennas, a fact that renders the proposed algorithm suitable for energy sustainable \textit{massive} MIMO systems considered in 5G applications.} Our detailed numerical investigation of the proposed joint TX and RX design validated the presented analysis and provided insights on the variation of the rate-energy trade off and the role of various system parameters. It was shown that our design results in nearly doubling the harvested power compared to benchmark schemes, thus enabling efficient MIMO SWIPT communication. \textcolor{black}{This trend holds true for any practical non-linear RF EH model.} We intend to extend our optimization framework in multiuser MIMO communication systems and consider the more general non-uniform PS reception in future works.

\appendices
\setcounter{equation}{0}
\setcounter{figure}{0}
\renewcommand{\theequation}{A.\arabic{equation}}
\renewcommand{\thefigure}{A.\arabic{figure}}
\section{Proof of Lemma~\ref{lem:GOP2}}\label{App:GOP2}  
We associate the Lagrange multipliers $\nu_2\geq0$ and $\mu_2\geq0$ with the constraints $({\rm C2})$ and $({\rm C5})$ in $\mathcal{OP}2$ while keeping $({\rm C3})$ implicit. The Lagrangian function for $\mathcal{OP}2$ can be written as
\begin{eqnarray}\label{eq:Lang2}
\mathcal{L}_2\left(\mathbf{S},\mu_2,\nu_2\right) \triangleq \mathrm{tr}\left(\mathbf{H}\mathbf{S}\mathbf{H}^{\rm H}\right)\!-\!\nu_2\left(\mathrm{tr}\left(\mathbf{S}\right)\!-\!P_T\right)\!
-\!\mu_2\left({R}\!-\!\log_2\left(\det\left(\mathbf{I}_{N_R}\!+\!\sigma^{-2}\mathbf{H}\mathbf{S}\mathbf{H}^{\rm H}\right)\right)\right).
\end{eqnarray}
Let us first investigate the $R>{R_{\mathrm{th}}^{\mathrm{id}}}$ scenario, where for fixed $\mu_2>0$ and $\nu_2>0$, the problem of finding $\mathbf{S}$ that maximizes the Lagrangian $\mathcal{L}_2\left(\mathbf{S},\mu_2,\nu_2\right)$ is expressed using \eqref{eq:Lang2} as 
\begin{equation*}\label{eqOPT3}
  \mathcal{OP}3: \max_{\mathbf{S}} \quad \log_2\left(\det\left(\mathbf{I}_{N_R}+\sigma^{-2}\mathbf{H}\mathbf{S}\mathbf{H}^{\rm H}\right)\right)-\mathrm{tr}\left(\mathbf{Q}\mathbf{S}\right)
	~~\textrm{s.t.}~~({\rm C3}),
\end{equation*}
where matrix $\mathbf{Q}\in\mathbb{C}^{N_T\times N_T}$ is defined as $\mathbf{Q}\triangleq \frac{\ln2}{\mu_2}\left(\nu_2\mathbf{I}_{N_T}-\mathbf{H}^{\rm H}\mathbf{H}\right)$. Problem $\mathcal{OP}3$ has a structure similar to the problem in \cite[eq. (16)]{MIMO_SWIPT} and its bounded optimal value can be obtained for arbitrary $\mathbf{Q}\succ0$, $\mu_2\geq0$, and $\nu_2>\lambda_{\max}\left(\mathbf{H}^{\rm H}\mathbf{H}\right)$ as
\begin{equation}\label{eq:optS2a}
\mathbf{S}_{\text{id}}^*\triangleq \mathbf{Q}^{-\frac{1}{2}} \Tilde{\mathbf{V}}\Tilde{\boldsymbol{\Lambda}}_{\rm o}\Tilde{\mathbf{V}}^{\rm H}\mathbf{Q}^{-\frac{1}{2}}
=\Tilde{\mathbf{V}}\left(\frac{\ln2}{\mu_2}\left(\nu_2\mathbf{I}_{r}-\boldsymbol{\Lambda}^{\rm H}\boldsymbol{\Lambda}\right)\right)^{-1}\Tilde{\boldsymbol{\Lambda}}_{\rm o}\Tilde{\mathbf{V}}^{\rm H},
\end{equation}
where unitary matrix $\Tilde{\mathbf{V}}\in\mathbb{C}^{N_T\times r}$ is obtained from the reduced SVD of the matrix $\sqrt{\sigma^{-2}}\mathbf{H}\mathbf{Q}^{-\frac{1}{2}}=\Tilde{\mathbf{U}}\Tilde{\boldsymbol{\Lambda}}\Tilde{\mathbf{V}}^{\rm H}$ with unitary matrix $\Tilde{\mathbf{U}}\in\mathbb{C}^{N_R\times r}$ and diagonal matrix $\Tilde{\boldsymbol{\Lambda}} \in\mathbb{C}^{r\times r}$ containing the $r$ eigenvalues of $\sqrt{\sigma^{-2}}\mathbf{H}\mathbf{Q}^{-\frac{1}{2}}$ in decreasing order. The entries of diagonal matrix $\Tilde{\boldsymbol{\Lambda}}_{\rm o}  \in\mathbb{C}^{r\times r}$,  obtained by using waterfilling solution~\cite{MIMO_book}, are related with the diagonal entries of $\Tilde{\boldsymbol{\Lambda}}$ as
\begin{equation}\label{eq:SWF}
[\Tilde{\boldsymbol{\Lambda}}_{\rm o}]_{i,i} =\left(1-[\Tilde{\boldsymbol{\Lambda}}]_{i,i}^{-2}\right)^+,\,\forall\,i=1,2,\ldots r.
\end{equation}
It is noted that the right-hand side of the equality in \eqref{eq:optS2a} results from rewriting $\mathbf{Q}$  using the reduced  SVD of $\mathbf{H}$ as $\mathbf{Q}=\mathbf{V}\left(\frac{\ln2}{\mu_2}\left(\nu_2\mathbf{I}_{r}-\boldsymbol{\Lambda}^{\rm H}\boldsymbol{\Lambda}\right)\right)\mathbf{V}^{\rm H}$, yielding
\begin{equation}\label{eq:simp1}
\sqrt{\sigma^{-2}}\mathbf{H}\mathbf{Q}^{-\frac{1}{2}}=\mathbf{U}\frac{\boldsymbol{\Lambda}}{\sqrt{\sigma^{2}}}\left(\frac{\ln2}{\mu_2}\left(\nu_2\mathbf{I}_{r}-\boldsymbol{\Lambda}^{\rm H}\boldsymbol{\Lambda}\right)\right)^{-\frac{1}{2}}\mathbf{V}^{\rm H}.
\end{equation}
Clearly, \eqref{eq:simp1} is the reduced SVD of matrix $\sqrt{\sigma^{-2}}\mathbf{H}\mathbf{Q}^{-\frac{1}{2}}$. Thus, we set $\Tilde{\mathbf{V}}=\mathbf{V}$, $\Tilde{\mathbf{U}}=\mathbf{U}$, and 
\begin{equation}\label{eq:simp2}
\Tilde{\boldsymbol{\Lambda}}= \frac{\boldsymbol{\Lambda}}{\sqrt{\sigma^{2}}}\left(\frac{\ln2}{\mu_2}\left(\nu_2\mathbf{I}_{r}-\boldsymbol{\Lambda}^{\rm H}\boldsymbol{\Lambda}\right)\right)^{-\frac{1}{2}}.
\end{equation}

Finally, we write $\mathbf{S}_{\text{id}}^*$ as $\mathbf{S}_{\text{id}}^* = \mathbf{F}_{\text{id}}\mathbf{F}_{\text{id}}^{\rm H}$ where $\mathbf{F}_{\text{id}} \triangleq \mathbf{V}\mathbf{P}_{\text{id}}^{1/2}$ with the diagonal matrix $\mathbf{P}_{\text{id}}$ defined as $\mathbf{P}_{\text{id}}\triangleq\Big(\frac{\ln2}{\mu_2}\big(\nu_2\mathbf{I}_{r}-$ $\boldsymbol{\Lambda}^{\rm H}\boldsymbol{\Lambda}\big)\Big)^{-1}\Tilde{\boldsymbol{\Lambda}}_{\rm o}$. Combining \eqref{eq:SWF} and \eqref{eq:simp2}, {the diagonal entries of $\mathbf{P}_{\text{id}}$ are} 
\begin{align}
p_k^{(\rm id)} = \left(\frac{\mu_2^*}{\ln2\left(\nu_2^*-[\boldsymbol{\Lambda}]_{k,k}^2\right)} -\frac{\sigma^2}{[\boldsymbol{\Lambda}]_{k,k}^2}\right)^+\quad\forall\,k=1,2,\ldots, r.
\end{align} 

For $R\le{R_{\mathrm{th}}^{\mathrm{id}}}$, $\mathbf{S}_{\text{id}}^*=\mathbf{S_{_\mathrm{EB}}}=P_T\,\mathbf{v}_1\mathbf{v}_1^{\rm H}$ is deduced from the discussion in Section~\ref{sec:tradeoff}. {Here} $({\rm C5})$ is satisfied at strict inequality and holds $\mu_2^*=0$ and $\nu_2^*=[\boldsymbol{\Lambda}]_{1,1}^2$. This completes the proof.

\bibliographystyle{IEEEtran}
\bibliography{refs_MIMO_PS_TBF}
\end{document}